\documentclass[10pt,aps,prd,twocolumn,nofootinbib,showpacs,showkeys,superscriptaddress]{revtex4-2}

\usepackage{psfrag} \usepackage{mathrsfs} \usepackage{amssymb, bm} 
\usepackage{amsmath, amsthm} \usepackage{epstopdf} 
\usepackage[breaklinks=true]{hyperref} \usepackage{enumerate} 
\usepackage{longtable} \usepackage{subfigure} \usepackage{color} 
\usepackage{mathrsfs} \usepackage{graphicx} 
\usepackage{bm,natbib,url,textcase}  \usepackage{xurl} 
\usepackage[title]{appendix}

\usepackage[english]{babel} \usepackage[T1]{fontenc} \usepackage{comment} 
\numberwithin{equation}{section}

\usepackage{geometry}
\allowdisplaybreaks

\hypersetup{
}

\usepackage{tikz}

\newcommand{\be}{\begin{equation}} 
\newcommand{\ee}{\end{equation}}

\definecolor{purple}{rgb}{1,0,1} \definecolor{lime}{HTML}{a6CE39} 

\newcommand{\orcidicon}{%
	\begin{tikzpicture}
		\draw[lime, fill=lime] (0,0) circle [radius=0.15] 
		node[white] {{\fontfamily{qag}\selectfont \tiny ID}}; 
		\draw[white, fill=white] (-0.0625,0.095) circle 
		[radius=0.007];
	\end{tikzpicture} \hspace{-2mm} }

\newcommand\orcidValerio{{\href{https://orcid.org/0000-0002-2601-1870}{\orcidicon}}}
\newcommand\orcidSantiago{{\href{https://orcid.org/0009-0001-8474-145X}{\orcidicon}}}

\begin{document} \def\theequation{\arabic{section}.\arabic{equation}}


\title{The thermal view of $f(R)$  cosmology}

\author{Valerio Faraoni\orcidValerio} \email[]{vfaraoni@ubishops.ca} 
\affiliation{Department of Physics \& Astronomy, Bishop's University, 2600 
College Street, Sherbrooke, Qu\'ebec, Canada J1M~1Z7}

\author{Santiago Novoa Cattivelli\orcidSantiago} 
\email[]{snovoa23@@ubishops.ca} 
\affiliation{Department of Physics \& Astronomy, Bishop's University, 2600 
College Street, Sherbrooke, Qu\'ebec, Canada J1M~1Z7}

\begin{abstract}

A new thermal view of scalar-tensor gravity, in which general relativity 
is the zero-temperature state of gravity, is applied to the 
subclass of $f(R)$ gravity theories and, specifically, to spatially 
homogeneous and isotropic universes. Within the limits of application of 
the new thermal formalism, results on the convergence to Einstein 
cosmology (or lack thereof) are first obtained for general $f(R)$ 
theories, and then illustrated with power-law and Starobinsky $f(R)$ 
gravity.

\end{abstract}


\maketitle \section{Introduction} 
\label{sec:1} 
\setcounter{equation}{0}

General Relativity (GR) has been very successful in the regimes where it 
is tested with good precision, but it cannot be the ultimate theory of 
gravity. If gravity and quantum mechanics have to merge at the Planck 
scale, as it is widely believed, GR must be changed already at low energy. 
The smallest quantum corrections modify the Einstein-Hilbert action by 
raising the order of the field equations and introducing extra propagating 
degrees of freedom \cite{Stelle:1976gc,Stelle:1977ry}.  Likewise, the 
low-energy limit of the bosonic string (the simplest string theory, which  
still incorporates essential features of string theories) 
does not reproduce GR but gives, instead, an $\omega=-1$ Brans-Dicke 
theory \cite{Callan:1985ia,Fradkin:1985ys} (where $\omega$ is the 
Brans-Dicke coupling constant introduced below).

The simplest way of adding a new degree of freedom to GR is by making Newton's 
constant $G$ a field varying across space and time, as proposed in the 
original Brans-Dicke theory \cite{Brans:1961sx} containing a gravitational 
scalar field $\phi$ and an effective gravitational coupling 
$G_\mathrm{eff} \simeq 1/\phi$. Brans-Dicke theory was later generalized 
to 
``first generation'' scalar-tensor gravity 
\cite{Bergmann:1968ve,Nordtvedt:1968qs,Wagoner:1970vr, Nordtvedt:1970uv}, 
still containing only a single extra scalar field $\phi$ in addition to 
GR, but making this scalar massive by endowing it with a potential 
$V(\phi)$, plus allowing the coefficient of the kinetic term of $\phi$ in 
the action to become a function of $\phi$ 
\cite{Bergmann:1968ve,Nordtvedt:1968qs,Wagoner:1970vr, Nordtvedt:1970uv}.

Enter cosmology: early universe inflation, although not proved, is widely 
regarded as a paradigm of modern cosmology. Starobinsky inflation 
\cite{Starobinsky}, which is based on quadratic corrections to the GR 
Lagrangian, stands as the most viable scenario of inflation and does not 
require the introduction of an {\em ad hoc} inflaton field, since the 
quadratic corrections that modify GR produce the same dynamics.

From a phenomenological point of view, the acceleration of the present-day 
cosmic expansion discovered with Type Ia supernovae in 1998 has prompted 
the introduction of a mysterious dark energy with exotic properties (see 
\cite{Amendola:2015ksp} for a review), and today there are (too) many 
theoretical models for this {\em ad hoc} dark energy, which looks more and 
more like a fudge factor. Moreover, the standard $\Lambda$-Cold Dark 
Matter 
($\Lambda$CDM) model of cosmology now exhibits worrysome tensions 
\cite{Riess:2019qba,DiValentino:2021izs}.  About twenty years ago, 
cosmologists resorted to alternative theories of gravity to avoid 
introducing the dark energy altogether. The idea is that, while gravity 
behaves as GR at smaller scales, it deviates from it at large, 
cosmological scales and the present acceleration of the cosmic expansion 
is due to these deviations. Among many proposals, the class of $f(R)$ 
theories of gravity has become very popular for modelling the present 
universe and a proof of principle has been given that this is possible, 
although these theories are not free of problems (see 
\cite{Sotiriou:2008rp,DeFelice:2010aj,Nojiri:2010wj} for reviews). 
$f(R)$ gravity is nothing but scalar-tensor gravity in disguise and adds a 
scalar degree of freedom $\phi=f'(R)$ to the two massless spin two degrees 
of freedom of GR \cite{Sotiriou:2008rp,DeFelice:2010aj, Nojiri:2010wj}.

In the last decade, the quest for the most general scalar-tensor gravity with 
equations of motion of only second order has led to rediscovering Horndeski gravity 
\cite{Horndeski:1974wa}, and to further generalize it with Degenerate Higher Order 
Scalar-Tensor (DHOST) theories. There is now a vast literature on scalar-tensor 
gravity and cosmology and it is difficult to gain a comprehensive view of all their 
aspects and observational constraints at different scales and regimes across the 
universe. A recent proposal (\cite{Faraoni:2018qdr, 
Faraoni:2021lfc,Faraoni:2021jri,Giusti:2021sku,Faraoni:2022gry, 
Giardino:2022sdv,Gallerani:2024gdy, Faraoni:2025alq,Faraoni:2025ufi}, see 
also \cite{Pereira:2025dmk}) views 
scalar-tensor gravity as a thermally excited state of GR, which corresponds to the 
``zero-temperature'' state of gravity. The approach of scalar-tensor gravity to GR 
occurring, for example, in the early universe \cite{Damour:1992kf,Damour:1993id} 
(with its complications \cite{Serna:2002fj}), is seen as the analogue of heat 
conduction in an effective dissipative fluid associated with the scalar degree of 
freedom $\phi$. A new formalism describing these situations (under certain 
restrictions) is currently under development and is being tested on various 
scalar-tensor theories and their exact solutions. Here we continue this work, 
applying this thermal picture (described below) to $f(R)$ cosmology. In this 
context, the new thermal view of scalar-tensor gravity unifies several theoretical 
results that appeared in a rather fragmented literature over the years and predicts 
when, and under what conditions, $f(R)$ universes converge to, or depart from, GR 
ones. Two basic ideas of the thermal view of scalar-tensor gravity are: i)~gravity 
is ``hot'' (i.e., deviates from GR) in strong gravity, in particular near spacetime 
singularities; ii)~the expansion of 3-space ``cools'' gravity, making it approach GR 
analogously to heat conduction in a dissipative fluid. Although these basic ideas 
prove true in simplified situations, more involved regimes involve the presence of 
``heat sources'' or ``sinks'' in the relevant equations, which complicate the basic 
picture. These more involved situations occur in $f(R)$ cosmology, as discussed in 
the following sections.

The new thermal formalism basically consists of an analogy between the 
dynamics of gravity and heat dissipation in a relativistic imperfect 
fluid. This analogy has limitations in its regime of applicability and, 
like all analogies, when using it one should not confuse its two sides, 
which describe physically different systems. However, since there are two 
sides to the 
analogy, one can use one of them (the most intuitive and better known 
one) to understand the other (less known) side, which is what we do 
here. 

To begin, let us introduce scalar-tensor gravity and its new thermal 
description.  We follow the notation of Ref.~\cite{Wald:1984rg}, using 
units in which the speed of light $c$ and Newton's constant $G$ are unity, 
and the metric signature is ${-}{+}{+}{+}$. First-generation scalar-tensor  
gravity is  described by the Jordan frame 
action \cite{Brans:1961sx,Bergmann:1968ve,Nordtvedt:1968qs,Wagoner:1970vr, 
Nordtvedt:1970uv}
\be 
S_\mathrm{ST} = \int \frac{d^4 x}{16\pi} \sqrt{-g} \left[ \phi R 
-\frac{\omega}{\phi} \, \nabla^c \phi \nabla_c \phi -V(\phi) \right] + 
S^\mathrm{(m)} \,, \label{action} 
\ee 
where $g$ is the determinant of the spacetime metric $g_{ab}$ with Ricci 
scalar $R$, $\phi>0$ is the Brans-Dicke-like scalar field 
(approximately corresponding to the inverse of the effective gravitational 
coupling 
$G_\mathrm{eff} \simeq 1/\phi$), $V(\phi)$ is the scalar field   
potential,  
the 
``Brans-Dicke coupling'' $\omega( \phi) >-3/2$ to avoid $\phi$ being  a 
phantom field, and $S^\mathrm{(m)}$ is the matter action.  The Jordan 
frame field equations read
\begin{eqnarray}  
&& R_{ab}-\frac{1}{2} \,  g_{ab}R = \frac{8\pi}{\phi} \, 
T_{ab}^\mathrm{(m)} + T_{ab}^{(\phi)} \,, \label{fe1}\\ 
&& \,\,\,T_{ab}^{(\phi)} = \frac{\omega}{\phi^2} \left( \nabla_a \phi 
\nabla_b 
\phi -\frac{1}{2} \, g_{ab} \nabla^c \phi \nabla_c \phi \right) 
\nonumber\\ 
&& \qquad\quad\quad +\frac{1}{\phi} \left( \nabla_a \nabla_b \phi - g_{ab} 
\Box\phi \right) -\frac{V}{2\phi} \, 
g_{ab} \, , \, \label{fe1.5}\\
&&\nonumber\\
&& (2\omega+3) \Box \phi = 8\pi T^\mathrm{(m)} 
+\phi \, V' -2V -\frac{d\omega}{d\phi}  \, \nabla^c\phi \nabla_c\phi \,, 
\nonumber\\  \label{fe2} 
\end{eqnarray} 
where $R_{ab}$ is the Ricci tensor, 
\begin{equation}
\label{set}
T_{ab}^\mathrm{(m)} =\frac{-2}{\sqrt{-g}}\, \frac{\delta
S^\mathrm{(m)} }{\delta g^{ab}}  
\end{equation}
is the matter 
stress-energy tensor, $T^{(m)} \equiv g^{ab} T_{ab}^\mathrm{(m)}$, 
$\nabla_a$ is 
the covariant derivative operator of $g_{ab}$, $\Box \equiv g^{ab} 
\nabla_a 
\nabla_b$, and $T_{ab}^{(\phi)}$ is an effective stress-energy tensor of 
$\phi$.  

Let us come now to the new thermal view of scalar-tensor gravity 
\cite{Faraoni:2018qdr, 
Faraoni:2021lfc,Faraoni:2021jri,Giusti:2021sku,Faraoni:2022gry,
Giardino:2022sdv,Gallerani:2024gdy, Faraoni:2025alq,Faraoni:2025ufi} (here  
we limit ourselves to ``first-generation'' scalar-tensor theories, which 
is all we need for $f(R)$ cosmology). The basic idea is to regard 
$T_{ab}^{(\phi)}$ in Eq.~(\ref{fe1.5}) as the effective stress-energy 
tensor of 
a fluid, which is possible when the gradient $\nabla^a \phi$ is timelike 
and future-oriented (a basic restriction of the formalism). Then, we 
define the effective fluid's four-velocity 
\be 
u^a \equiv \frac{ \nabla^a\phi}{\sqrt{ -\nabla_c\phi \nabla^c\phi}} 
\ee
and $T_{ab}^{(\phi)}$ assumes the structure of a dissipative fluid 
stress-energy tensor \cite{Eckart:1940te,Maartens:1996vi,Andersson:2006nr} 
\be
T_{ab}^{(\phi)} =\rho^{(\phi)}  u_a u_b  + P^{(\phi)}   h_{ab} 
+\pi_{ab}^{(\phi)}   +q_a^{(\phi)}  u_b +q_b^{(\phi)}  u_a \,,
\ee
where $h_{ab} \equiv g_{ab}+u_a u_b$ is the Riemannian metric on the 
3-space experienced by observers comoving with the fluid, 
\begin{eqnarray}
\rho^{(\phi)} =  -\frac{\omega \, \nabla^e \phi \nabla_e 
\phi}{2\phi^2}   +  \frac{V}{2\phi} + 
\frac{1}{\phi} \left( \square \phi -  
\frac{  \nabla^a \phi \nabla^b \phi \nabla_a 
\nabla_b \phi}{ \nabla^e \phi  \nabla_e \phi  } \right) \nonumber\\
\label{effdensity}
\end{eqnarray}
is an effective energy density,
\be
P^{(\phi)}  =  - \frac{\omega\, \nabla^e \phi \nabla_e \phi }{2\phi^2} 
 -  \frac{V}{2\phi} - \frac{1}{3\phi}  \left( 2\square \phi + 
\frac{\nabla^a \phi \nabla^b \phi \nabla_b \nabla_a \phi }{\nabla^e \phi 
\nabla_e  \phi }  \right)  \label{effpressure}
\ee
is an effective isotropic pressure,
\begin{widetext}
\begin{eqnarray}
\pi_{ab}^{(\phi)}   &=& \frac{1}{\phi \nabla^e \phi \nabla_e 
\phi } 
\left[ \frac{1}{3} \left( \nabla_a  \phi \nabla_b \phi - g_{ab} \nabla^c 
\phi \nabla_c \phi \right) \left(  \square \phi  - 
\frac{ \nabla^c \phi  \nabla^d \phi \nabla_d \nabla_c \phi }{ \nabla^e 
\phi \nabla_e \phi }   
\right) \right. \nonumber\\
&&\nonumber\\
&\, & \left. + \nabla^d \phi \left(  \nabla_d \phi \nabla_a \nabla_b 
\phi - 
\nabla_b \phi \nabla_a \nabla_d  \phi - \nabla_a \phi \nabla_d \nabla_b 
\phi +  
\frac{ \nabla_a \phi \nabla_b \phi  \nabla^c \phi \nabla_c 
\nabla_d \phi }{ \nabla^e \phi \nabla_e \phi } \right) \right] 
\label{piab-phi}
\end{eqnarray}
\end{widetext} 
is the (purely spatial and trace-free) stress tensor, while  
\be
 q_a^{(\phi)} = \frac{\nabla^c  \phi \nabla^d \phi}{\phi 
  \left(-\nabla^e \phi \nabla_e \phi \right)^{3/2} } \,  
\Big(  \nabla_d \phi \nabla_c \nabla_a \phi 
- \nabla_a \phi \nabla_c \nabla_d \phi \Big)  \label{eq:q}
\ee
is an effective heat flux density \cite{Faraoni:2018qdr}.   The second 
derivatives of $\phi$ endow this effective fluid with a 
dissipative 
character. This structure is 
common to all symmetric two-index tensors and there is no physics in it 
\cite{Faraoni:2023hwu}, but then a 
little miracle happens. Eckart's theory of dissipative fluids 
\cite{Eckart:1940te,Maartens:1996vi,Andersson:2006nr}  
assumes three constitutive relations, the most important of which is the 
generalized Fourier law
\be
q_a = - {\cal K} h_{ab} \left( \nabla^b {\cal T}+ {\cal T} \dot{u}^b 
\right) 
\,,
\ee
where ${\cal K}$ is the thermal conductivity of the fluid, ${\cal T}$ is 
its  
temperature, and $\dot{u}^a \equiv u^c \nabla_c u^a$ is the fluid  
four-acceleration. A direct computation \cite{Faraoni:2018qdr} shows that 
$q_a^{(\phi)}$ is proportional to $\dot{u}_a$, which allows one to 
identify its coefficient with 
\be  
{\cal K}{\cal T} = \frac{  \sqrt{-\nabla^c\phi \nabla_c\phi} }{ 8\pi \phi} 
\,, \label{KTdefinition} 
\ee 
thus defining the product of the ``effective thermal conductivity'' and 
the ``effective temperature of gravity''. GR is obtained for 
$\phi=$~const. and corresponds to zero temperature, the state of thermal 
equilibrium. Regimes in which the scalar degree of freedom is excited and 
propagates correspond to ``warmer'' states at ${\cal K}{\cal T}>0$ 
\cite{Faraoni:2021lfc,Faraoni:2021jri,Giusti:2021sku,Faraoni:2022gry,
Giardino:2022sdv,Gallerani:2024gdy, Faraoni:2025alq,Faraoni:2025ufi}. 
A number of results have been obtained in the formalism, including 
the evolution equation for ${\cal K}{\cal T}$ 
\cite{Faraoni:2021lfc,Faraoni:2021jri,Giusti:2021sku,Faraoni:2022gry} 
\begin{eqnarray}
\frac{d \left( {\cal K}{\cal T}\right)}{d\tau} &=& 8\pi \left( {\cal 
K}{\cal T}\right)^2 
-\Theta \, {\cal K}{\cal T} + \frac{ T^\mathrm{(m)} }{\left(  2\omega + 3 
\right) 
\phi} \nonumber\\
&&\nonumber\\
&\, &  +\frac{1}{8\pi 
\left( 2\omega + 3 \right)} \left(  V' -\frac{2V}{\phi} 
-\frac{1}{\phi} \, \frac{d \omega}{d\phi} \, 
\nabla^c\phi \nabla_c\phi \right) \,,\nonumber\\
&& \label{evolution_general2} 
\end{eqnarray}
where $\tau$ is the proper time along the effective $\phi$-fluid lines 
and $\Theta \equiv \nabla_c u^c$ is its expansion scalar.

Gravity is ``heated'' (i.e., $d\left( {\cal K}{\cal T}\right)/d\tau>0$) by 
positive terms in the right-hand side and is ``cooled'' (i.e., $d\left( 
{\cal K}{\cal T} \right)/d\tau<0$) by negative ones.

The recent  Ref.~\cite{Faraoni:2025alq} advances the thermal picture with 
the study of scalar-tensor situations in which $\Box\phi=0$, in which case 
the $\left( \Theta, {\cal K}{\cal T} \right)$ plane (not a phase 
space) provides  a convenient 
representation of the behavior of gravity. If the expansion scalar 
$\Theta $ is negative, 
or if the system begins at a point above the critical half-line $ 8\pi 
{\cal K}{\cal T}= \Theta>0$, gravity either ``heats up'' diverging away 
from GR, or else this line is crossed with horizontal tangent, after 
which gravity cools forever \cite{Faraoni:2025alq,Faraoni:2025dex}.

 If instead $ \Theta>0$ and 
the system begins at a point below this critical half-line (i.e., with  $ 
8\pi {\cal K}{\cal T}<\Theta$), gravity can only ``cool'' and converge to 
GR \cite{Faraoni:2025alq}. This picture survives if conformal matter with 
$T^{(m)}=0$, but no potential, is added to the picture. Here we study 
vacuum $f(R)$ gravity, which unavoidably adds a potential for  
$\phi=f'(R)$. 

Section~\ref{sec:2} reviews $f(R)$ gravity and cosmology, while the 
following sections apply the thermal view to it. Several exact solutions of $f(R)$ 
gravity are known 
(e.g., 
\cite{Cognola:2005de,Nojiri:2006gh,Capozziello:2006dj,Cognola:2008zp,Nojiri:2009kx}) 
and we use some of them as examples.

\section{$f(R)$ gravity } 
\label{sec:2} 
\setcounter{equation}{0}

We focus on a subclass of scalar-tensor gravities, i.e., metric 
$f(R)$ gravity described by the action
\be
S= \frac{1}{16\pi } \int d^4 x \sqrt{-g} \, f(R) +S^\mathrm{(m)} 
\,,\label{f(R)action}
\ee
where $f(R)$ is a non-linear function of  the Ricci scalar. The 
fourth order field equations produced by varying the 
action~(\ref{f(R)action}) with respect to the inverse metric $g^{ab}$  are 
\cite{Sotiriou:2008rp,DeFelice:2010aj, Nojiri:2010wj}
\be
 f'(R)R_{ab}-\frac{f(R)}{2} \, g_{ab} = 
8\pi  \,T_{ab}^\mathrm{(m)} 
+ \left( \nabla_a \nabla_b -g_{ab}\Box\right) f'(R)
\,,  \label{metf}
\ee
where a prime denotes differentiation with respect to the argument. 
Tracing Eq.~(\ref{metf}) yields 
 \be
\label{metftrace}
\Box f'  + \frac{1}{3} \left[ f'(R)R-2f(R) \right] =\frac{8\pi}{3} 
\,T^\mathrm{(m)}  \,,
\ee
which implies that $f'(R)$ is a dynamical, propagating degree of 
freedom.

It is well 
known \cite{Sotiriou:2008rp,DeFelice:2010aj, Nojiri:2010wj} 
that metric $f(R)$ gravity is equivalent to a Brans-Dicke theory with 
scalar field $\phi =f'(R)$, Brans-Dicke parameter $\omega=0$, and scalar 
field potential
\be
V(\phi) = \phi R -f(R) \Bigg|_{ f'(R)=\phi} \,.\label{potential}
\ee
In general, this potential cannot be written explicitly in terms of 
$\phi$.

In $f(R)$ gravity, it must be $ f'(R)>0$ in order for the gravitational 
coupling to be positive and the graviton to carry positive kinetic energy. 
Additionally, it must be $f''(R)>0$ to avoid local tachyonic instabilities 
\cite{Dolgov:2003px,Faraoni:2006sy}.

In the Friedmann-Lema\^itre-Robertson-Walker (FLRW) universe with line 
element in comoving polar coordinates $\left( t,r, \vartheta, \varphi \right)$
\be
ds^2=-dt^2 + a^2(t) \left[ \frac{dr^2}{1-kr^2}+r^2 \left( d\vartheta^2 + 
\sin^2\vartheta \, d\varphi^2 \right)\right] 
\ee
(where $k=0, \pm 1$ is the normalized curvature index), the $f(R)$ field 
equations assume the form
\begin{eqnarray}
\label{H_squa}
&& H^2  =  \frac{1}{3f'}\left[ 8\pi\rho+ \frac{Rf'-f}{2}-3H\dot{R}f''\right] 
-\frac{k}{a^2}  \,,\nonumber\\
&&\\
&& 2\dot{H} +3H^2  =  -\frac{1}{f'}\Big[ 8\pi P + (\dot{R})^2 f'''  
+2H\dot{R}f''  \nonumber\\
&& +\ddot{R}f''+\frac{1}{2}\left( f-Rf' \right)\Big] \,,  \label{H_dot} 
\end{eqnarray}
where $\rho$ and $P$ are the energy density and pressure of the cosmic 
matter fluid, respectively. 

The evolution equation~(\ref{evolution_general2})  for ${\cal K}{\cal T}$ 
in electrovacuum $f(R)$ gravity reads
\be
\frac{ d\left( {\cal K}{\cal T}\right)}{d\tau} = {\cal K}{\cal T}\left( 
8\pi 
{\cal K}{\cal T}-\Theta\right) + \frac{ 2f(R)-Rf'(R)}{24 \pi f'(R) } \,. 
\label{KTf(R)}   
\ee

In a FLRW universe, the gradient $\nabla^a \phi$ of the Brans-Dicke-like 
scalar field $\phi(t) =f'(R(t))$ is always timelike. Denoting with an 
overdot the differentiation with respect to the comoving time $t$, we have
\be
\dot{\phi} = f''(R) \dot{R} \label{zzz}
\ee
and $\dot{\phi} $ has the sign of $\dot{R}$ since we require $f''(R) >0$. 
In order for 
\be
 u^a= \frac{ \nabla^a\phi}{\sqrt{-\nabla^c\phi \nabla_c\phi } } = 
\frac{\nabla^aR}{\sqrt{-\nabla^cR \nabla_cR}} 
\ee  
to be future-oriented, it must be 
$\dot{\phi}<0$ because
\be
\nabla^a\phi = g^{ab} \nabla_b \phi = g^{ab} \, {\delta_b}^0 \, \partial_t 
\phi = -\dot{\phi} \, {\delta^a}_0 
\ee
and the time component $\nabla^0 \phi = - \dot{\phi}$ must be positive for 
the effective $\phi$-fluid of our formalism to have $u^0>0$ and evolve 
toward the future.  Therefore, according to Eq.~(\ref{zzz}), the thermal 
analogy is only applicable to FLRW solutions of $f(R)$ gravity for which 
\be
R(t)=6\left( \frac{ \ddot{a} }{a} + \frac{\dot{a}^2 }{ a^2} +\frac{k}{a^2} 
\right) \label{Ricciscalar}
\ee
{\it decreases} with time.

In a FLRW universe, the proper  time $\tau$ of the effective 
$\phi$-fluid 
coincides with the comoving time $t$. In fact, the four-velocity of the 
effective fluid has components  
\be
u^a \equiv  \frac{ \nabla^a \phi}{ \sqrt{ -\nabla^c\phi\nabla_c\phi} } = 
\frac{ -\dot{\phi} \, {\delta^a}_0 }{|\dot{\phi}| } ={\delta^a}_0 
\ee
in coordinates comoving with the effective $\phi$-fluid, 
therefore, 
\be
u^0 = \frac{dx^0}{d\tau} = \frac{dt}{d\tau}=1
\ee
and $t=\tau$ up to an irrelevant additive constant corresponding to the 
choice of the origin on the time axis, so time evolution with respect to 
$t$ coincides with time evolution with respect to $\tau$.

In FLRW universes we have 
\be
{\cal K}{\cal T} = \frac{ |\dot{\phi}|}{ 8\pi \phi} = \frac{ f''(R) 
|\dot{R}|}{8\pi f'(R)} \,.
\ee
All solutions with $R=$~const. (when they exist) are automatically states 
of thermal equilibrium at ${\cal K}{\cal T}=0$ \cite{Giardino:2023qlu}. 
They 
include the radiation era with $R=0$ discussed in the next section. 

Before proceeding to analyze the thermal view of $f(R)$ gravity, let us 
take a look at the phase space of spatially flat FLRW cosmology in these 
theories. The phase space of vacuum $ k=0$ scalar-tensor cosmology has 
been 
described in \cite{Faraoni:2005vc}. When $k=0$ (the universe preferred by 
observations), the scale factor $a(t)$ 
enters the field equations~(\ref{H_squa}), (\ref{H_dot}) only through the 
Hubble function $H\equiv 
\dot{a}/a$, which is a cosmological observable and is adopted as one of 
the phase space variables. The other variables are the scalar field $\phi$ 
and its time derivative $\dot{\phi}$. The phase space is, therefore, the 
3D space $\left( H, \phi, \dot{\phi} \right)$, but the Hamiltonian 
constraint (a first order equation similar to an energy integral) forces 
the orbits of the solutions of the field equations to move on a 2D subset  
of this 3D space. The Hamiltonian constraint for vacuum, spatially flat, 
FLRW universes in Brans-Dicke gravity has the form \cite{Faraoni:2005vc}
\be
H^2 = -H\, \frac{ \dot{\phi} }{\phi}  + \frac{\omega}{6}  \, \left( 
\frac{\dot{\phi} }{\phi} \right)^2 + \frac{V(\phi)}{6\phi} \,\,;
\ee
solving this formal quadratic equation for 
$\dot{\phi}$ in terms of $H$ and $\phi$ yields \cite{Faraoni:2005vc} 
\be
\dot{\phi} \left( H, \phi \right) = \frac{3\phi}{\omega} \left[ H  \pm 
\sqrt{ H^2 -\frac{2\omega}{3} \left( \frac{V}{6\phi} - H^2\right) } 
\, \right]  \,. \label{boh} 
\ee
In general, the plus or minus sign corresponds to two ``sheets'' in the 3D 
space, joining each other on the boundary of a region ${\cal F}$ forbidden 
to the trajectories of the dynamical system. This forbidden region 
corresponds to a negative argument of the square root in Eq.~(\ref{boh}), 
and its boundary $\partial {\cal F}$ to vanishing square root argument 
\cite{Faraoni:2005vc}.

The structure of the 3D phase space of vacuum, $k=0$ $f(R)$ cosmology is 
simpler, and was discussed in \cite{deSouza:2007zpn}. Again, $a(t)$ enters 
the field equations only through $H$, and we choose the 
cosmological observable $\Theta=3H$ as one of 
the phase space variables. Since $f'(R)>0$, there is a one-to-one 
correspondence between $R$ and $f(R)$ and we can choose the Ricci scalar 
$R$ as the second phase space variable.\footnote{$R$ and $\Theta$ are 
independent because $R$ depends on both $\Theta $ and $\dot{\Theta}$ 
according to Eq.~(\ref{Ricciscalar}).}
 Then, the third variable would be 
its derivative $\dot{R}$, but we can trade it for ${\cal K}{\cal T}=\frac{
|\dot{\phi}|}{8\pi \phi} = \frac{f''(R) |\dot{R}|}{8\pi f'(R)} $, limiting
ourselves to the dynamical situations in which $\dot{R}<0$ to be able to 
apply the thermal formalism. Since $f(R)$ gravity corresponds to an 
$\omega=0$ Brans-Dicke theory (with a potential), the quadratic term in 
$\dot{\phi}/\phi$ is absent from the Hamiltonian constraint, which 
formally reduces to a linear equation, enabling us to eliminate $\dot{R}$ 
from the field equations using  the single-valued function 
\cite{deSouza:2007zpn}
\begin{eqnarray} 
\dot{R}\left( \Theta, R \right) &=& \frac{Rf'(R)- f(R)- 6H^2 
f'(R)}{6H f''(R)} \nonumber\\
&&\nonumber\\
&=& \frac{Rf'(R)- f(R)- 2\Theta^2 f'(R)/3}{2\Theta f''(R)}  \,.
\end{eqnarray}
Therefore, the orbits of the solutions are forced to live on the 2D subset 
of the 3D phase space $\left( \Theta, R,  {\cal K}{\cal T} 
\right)$ described by 
\be
{\cal K}{\cal T} \left( \Theta, R \right) = \frac{1}{16 \pi f'(R)} \,
\left|  \frac{ Rf'(R)-f(R) -2\Theta^2 f'(R)/3}{\Theta} \right| \,.
\ee
The projections of the orbits on the $\left( \Theta, R \right)$ 
plane cannot intersect, contrary to more general scalar-tensor theories 
in which two sheets project onto this plane \cite{deSouza:2007zpn}. 

Depending on the form of the function $f(R)$, there may be a forbidden 
region ${\cal F}$ of the phase space corresponding to $H^2<0$ or (using 
Eq.~(\ref{H_squa}) with $\rho=0$ and the expressions of $\Theta$ and 
${\cal K}{\cal T}$),
\be
Rf'(R)-f(R)+ 16\pi \Theta f'(R) {\cal K}{\cal T} <0 \,.
\ee
 
Having chosen $\left( \Theta, R,  {\cal K}{\cal T} 
\right)$ as phase space  variables, if fixed points of the dynamical 
system exist, they correspond to $ H=$~const.$\equiv H_0$ and 
$R=$~const.$\equiv R_0 =12H_0^2 $, more 
precisely \cite{Barrow:1983rx,deSouza:2007zpn}
\be
\left( H, R, {\cal K}{\cal T} \right)= \left( \pm \sqrt{ \frac{f_0}{ 
6f_0'}}, \frac{2f_0}{f_0'}, 0 \right) 
\ee
(where $f_0 \equiv f(R_0)$ and $ f'_0\equiv f'(R_0)$). 
Since $k=0$, these fixed points can only be de Sitter spaces, possibly 
degenerating into the Minkowski point $\left( \Theta, R , {\cal K}{\cal T} 
\right)= \left( 0,0,0 \right)$. A radiation era with $R\equiv 0$ implies 
that also ${\cal K}{\cal T} \propto |\dot{R}| =0$ and its trajectory 
necessarily unfolds along the $\Theta$-axis, and is not a fixed point. 

The 2D sheet on which the orbits live separates an ``upper'' and a 
``lower'' region corresponding to $k=+1$ and $k=-1$ (for $ k\neq 0$, the 
orbits unfold in three phase space dimensions). These orbits cannot cross 
the $k=0$ ``sheet'' because the topology of 3-space cannot change 
dynamically.

\section{Thermal view of FLRW  cosmology in  general $f(R)$ gravity} 
\label{sec:3} 
\setcounter{equation}{0}

Several results about the convergence of $f(R)$ gravity to GR (i.e., 
${\cal K}{\cal T}\to0$) or its departure from it (i.e., increasing ${\cal 
K}{\cal T}$) can be obtained for general functions $f(R)$. To begin with, 
the authors of \cite{CotsakisFlessas1995} investigated the 
structural stability of the FLRW solutions 
of polynomial $f(R)$ gravity with respect to modifications of the 
field equations (from the Einstein equations to the fourth order 
equations~(\ref{metf}) of polynomial $f(R)$ gravity). They studied early 
times near the 
initial singularity $t\to 0^{+}$  and late times $t\to 
+\infty$, assuming the matter source to be a perfect fluid with constant 
equation of state $P=w\rho$, with equation of state parameter in the  
range $-1 \leq w\leq 1$ (in particular for 
radiation $w=1/3$), and for all values  of the curvature index 
$k$. This procedure makes sense only when the unperturbed solution is 
simultaneously a solution of GR and of $f(R)$ gravity. For general 
spacetimes, this is a very strong requirement and admits only a very 
restricted class of solutions \cite{Hervik:2017sdr,Hervik:2013cla}. 
However, any viable theory of gravity should admit FLRW solutions, which 
are indeed very common. The structural stability of FLRW solutions of GR 
when the Einstein-Friedmann equations are modified to the fourth-order 
equations of $f(R)$ gravity was previously  investigated by Barrow and 
Ottewill in the influential Ref.~\cite{Barrow:1983rx}.  A notable result 
of this reference  is that the FLRW solution for the radiation 
era solves exactly the fourth order equations of motion {\it for any 
$f(R)$ gravity with $f(0)=0$ and $f'(0) \neq 0$} (but only $f'(0)>0$ is 
physically admissible).

To understand this result, which has remained  a bit cryptic over 
the  years, note that the class of $f(R)$ theories with 
$f(0)=0$ and $f'(0)>0$ includes GR and its  radiation solutions, which are 
(e.g., \cite{Faraoni:2021opj}) 
\be
 a(t) = \left\{ \begin{array}{ccc}
 \sqrt{  \frac{8\pi}{3} \, \rho_\mathrm{r}^{(0)} -\left( 
t-t_0\right)^2 }   &  & \quad \mbox{if} \;\;  k=1  \,,\\
&&\\
a_0  \sqrt{t} & & \quad \mbox{if} \;\;  k=0 \,,\\
&&\\
 \sqrt{  \left( t-t_0\right)^2 - 
\frac{8\pi}{3} \, \rho_\mathrm{r}^{(0)} 
}   &  &  \quad \mbox{if} \;\;  k=-1 \,, \end{array} \right. 
\label{radiationsolutions} 
\ee
where the constant $\rho_\mathrm{r}^{(0)}$ is such that  the energy 
density of radiation is $\rho_\mathrm{r}(a) = 
\rho_\mathrm{r}^{(0)} /a^4$. It is easy to see, using 
Eq.~(\ref{Ricciscalar}), that these solutions  have Ricci scalar $R = 0$ 
regardless of the form of the function $f(R)$, a 
feature that persists in all these 
$f(R)$ theories. In this radiation era, $R$ remains zero 
all the time and  $\phi=f'(R)=f'(0)>0 $ remains constant, that is, the 
theory 
is forced to be  GR. From the thermal point of view, we have that ${\cal 
K}{\cal T}=0$ at all 
times {\em for all these $f(R)$ theories}, hence the structural stability 
of the solutions corresponds to the statement that the thermal state of 
equilibrium ${\cal K}{\cal T}=0$ is universal to these theories because 
$R$ 
is constant and the scalar degree of freedom $\phi=f'(R)$ is 
simply eliminated from them. Time evolution in the $\left( \Theta, {\cal 
K}{\cal T} \right) $ plane is simple: 

\begin{itemize}

\item For $k=0$, $\Theta 
=\frac{3}{2t} $  is infinite at the Big Bang $t=0$ and vanishes as $ t\to 
+ \infty$. Time evolution is represented by  motion strictly along the 
$\Theta$-axis, from $\Theta=+\infty$ towards $\Theta=0$.   

\item For $k=-1$, the universe begins with a Big Bang at $\Theta=+\infty$ 
and evolves towards the late-time $k=0$ solution with $\Theta=0$: the 
evolution in the $\left( \Theta, {\cal K}{\cal T}\right)$ plane is the 
same as in 
the previous case.

\item For $k=+1$, the universe begins with a Big Bang at 
$\Theta=+\infty$, expands to a maximum size where $\Theta=0$, 
and then contracts ($\Theta<0$) and ends in a  Big Crunch at  
$\Theta\to -\infty$. Time evolution consists of motion along the ${\cal 
K}{\cal T}=0$ axis from  $\Theta=+\infty $ to $\Theta=-\infty$. 

\end{itemize}

Gravity never moves away from GR even though $\Theta$ becomes negative 
because it is always forced to stay on the $\Theta$-axis and to be GR, and 
the effective $\phi$-fluid disappears. In fact, the combination  
$\left( \nabla_a\nabla_b - g_{ab}\Box \right)\phi$ in Eq.~(\ref{metf}) 
vanishes identically and this set of equations reduces to the Einstein 
equations.  

From both the dynamical and the thermal points of view, as long as we 
restrict to radiation universes, these $f(R)$ theories are not interesting 
alternatives to GR because no propagating degree of freedom is added to 
the two (massless) tensor modes of GR, and there is never the possibility 
of gravity moving away from GR. Equation~(\ref{KTf(R)}) is 
satisfied identically by ${\cal K}{\cal T}=0$. 

Even with the restriction $f(0)=0$, the radiation 
solution~(\ref{radiationsolutions}) is not, in general, the only solution 
of the fourth order equations of $f(R)$ gravity and other solutions are 
possible (see the case of Starobinsky gravity below for an example).

The authors of \cite{CotsakisFlessas1995} performed a perturbation 
analysis of GR solutions and found that in most (but not all)  
situations, the solutions of polynomial $f(R)$ gravity depart from those 
of GR (or, in their language are ``non-perturbative''). They are singular 
solutions different from the GR universe with the same fluid content. This 
means that, in the $f(R)$ theories considered, gravity is significantly 
different from GR as $t\to 0^{+}$ unless special restrictions are imposed 
on the coefficients of the independent modes in linear perturbation 
theory \cite{CotsakisFlessas1995}.

The authors of \cite{CotsakisFlessas1995} even conjectured that all 
``physically resonable'' 
FLRW solutions of GR are unstable against structural perturbations of the 
field equations that change them from the Einstein equations to the field 
equations of higher-order gravity \cite{CotsakisFlessas1995} (the 
radiation-dominated solutions of Barrow and Ottewill \cite{Barrow:1983rx} 
already described constitute an obvious exception).

We can frame this conjecture within the thermal view of $f(R)$ gravity, 
using the recent results of \cite{Faraoni:2025alq}. The discussion is not 
limited to polynomial or other forms of the function $f(R)$, but is 
general.

First, it is crucial that the dynamics near a Big Bang singularity is 
dominated by gravity instead of matter. This behavior is quite plausible 
\cite{Ruban:1972bg,Faraoni:2025ufi}, but is not a given and should be 
investigated case by case. When it turns out to be true, the dynamics 
reduces to that of vacuum $f(R)$ gravity. 

We now know that, for vacuum or conformal matter, whether the conjecture 
of \cite{CotsakisFlessas1995} is satisfied or not depends on the initial 
conditions.  Expanding FLRW universes which start out with $ 8\pi{\cal 
K}{\cal 
T}>\Theta>0$, plus contracting ones with $\Theta<0$,   
depart 
from GR, but solutions that start out with $8\pi {\cal K}{\cal T}<\Theta$ 
do approach GR \cite{Faraoni:2025alq}.  
Solutions with $8\pi {\cal 
K}{\cal  T}=\Theta$  and with constant 
${\cal K}{\cal 
T}$ and constant 
$\Theta$ are fixed points, but they are thermally unstable  
\cite{Faraoni:2025alq}.  The 
thermal picture settles the conjecture of \cite{CotsakisFlessas1995} for 
vacuum or conformal matter with a completely different approach. For more 
general forms of matter, for which the term $ \frac{ T^\mathrm{(m)} 
}{(2\omega+3)\phi} $ in Eq.~(\ref{evolution_general2}) is non-vanishing, 
the thermal picture is more complicated.

In their perturbative analysis, the authors of 
\cite{Ruzmaikina70,Cotsakis:1993zz,CotsakisFlessas1995} find 
$k=0$ universes that are regular at $t=0$ (contrary to GR universes that 
have a Big Bang singularity) and 
converge to GR solutions as $t\to +\infty$, apparently contradicting the 
general thermal view of scalar-tensor gravity. The latter  states that 
universes 
starting out with $8\pi {\cal K}{\cal T}>\Theta>0$ (which they  do if 
they initially depart drastically from GR)  run away from GR as long as 
$8\pi {\cal 
K}{\cal T}$ remains larger than $\Theta$ \cite{Faraoni:2025alq}.   
However, these universes of 
\cite{Ruzmaikina70,Cotsakis:1993zz,CotsakisFlessas1995} are obtained for 
negative values of the parameter 
$\lambda^2 \equiv \frac{f_0'}{3f_0''} $ (the zero subscript denotes 
quantities evaluated on the unperturbed solution).   
Unbeknownst to the 
authors of Refs.~\cite{Ruzmaikina70,Cotsakis:1993zz,CotsakisFlessas1995} 
working in the 1970s and 1990s, we now know that in $f(R)$ gravity it must 
be $ 
f'(R)>0$ and $f''(R)>0$ \cite{Sotiriou:2008rp,DeFelice:2010aj, 
Nojiri:2010wj}, therefore, the region $\lambda^2<0$ is physically 
irrelevant.\footnote{The same conclusion applies to the solutions~(13), 
(15), (19), (21), and~(27)  of Ref.~\cite{CotsakisFlessas1995}.}

To continue the discussion, let us apply more  results of 
Ref.~\cite{Faraoni:2025alq} to 
the $f(R)$ subclass of Brans-Dicke theory. Gravity ``heats up'' if the 
effective potential~(\ref{potential})  grows 
faster than $\sim \phi^2$ with $\phi=f'(R)$, which is expressed by the 
asymptotic condition
\be
V(\phi)= Rf'(R) - f(R)  > \beta \left[ f'(R)\right]^2 \,, 
\ee
where $\beta $ is a positive constant. 
Using the modified notation $x\equiv R$ and $ y(x) \equiv f(R)$, the non-linear 
ordinary differential equation setting the boundary of 
this situation,
\be
\beta y'^2 -xy' +y=0 \,,
\ee
admits the two linear solutions
\be
y_1(x) = \frac{c_1}{2\beta} \, x -\frac{c_1^2}{4\beta} \,, \quad \quad 
y_2(x) = -\frac{c_1}{4\beta} \, x -\frac{c_1^2}{16\beta} \,,
\label{y-solutions}
\ee
both of which give the boundary choice $f(R)=R-2\Lambda$ (i.e., GR) where, 
interestingly, only $\Lambda \geq 0$ is possible. This condition forbids, 
for example, Anti-de Sitter and asymptotically Anti-de Sitter universes. 
Hence, gravity ``heats up'' if $V(\phi)$ grows faster than $\phi^2 $ or
\be
y'\left( \beta y'-x \right)<-y <0 \,,
\ee
where $y>0$ and $y'>0$. This condition is satisfied for $0<y'<x/\beta $ 
(assuming $x \equiv R>0$), or $f'(R)< R/\beta$, which means that $f(R)$ 
grows slower than $R^2/\left( 2\beta \right)$. This condition {\it on the 
derivative} $f'(R)$ can be translated into a condition {\it on the 
function} $f(R)$ if we require that $f(0)=0$. This condition (required in 
the Barrow-Ottewill theorem already discussed \cite{Barrow:1983rx}) is 
satisfied by most physically relevant models in the literature, including 
$R^n$ and polynomial $f(R)$ gravity, as well as in popular 
$f(R)$ models aiming at explaining the 
cosmic acceleration without dark energy, such as the Hu-Sawicki scenario  
\cite{Hu:2007nk}
\be
f_\mathrm{HS}(R) = R- \frac{c_1 m^2 \left( R/m^2\right)^n }{ c_2 
\left( R/m^2  \right)^n +1 }  
\ee
(where $c_{1,2}$, $m^2$, and $n$ are constants), 
the Starobinsky model\footnote{Not to be confused with Starobinsky 
gravity $f(R)=R+\alpha R^2$.} \cite{Starobinsky:2007hu}
\be
f_\mathrm{S}(R) = R+\lambda R_s \left[ \left(1+\frac{R^2}{R_s^2} 
\right)^{-q} -1 
\right]  
\ee
(with parameters $\lambda, R_s, q$), and the Miranda-Joras-Waga-Quartin
 model \cite{Miranda:2009rs} 
\be  
f_\mathrm{MJWQ} (R) = R- \alpha_M R_{*} \ln \left( 1+\frac{R}{R_{*} } 
\right)   
\ee
(with two parameters  $\alpha_M, R_{*}$). 

The condition $f(0)=0$ anchors all the relevant functions to the 
origin of the $\left( R, f(R) \right)$ plane and allows us to integrate 
the inequality $f'(R)<R/\beta$ to $f(R) <R^2/(2\beta)$. 

We conclude that:

\begin{itemize}

\item gravity ``heats up'' if $f(R)$ grows slower than $R^2$ when $R$ 
increases (since we are limited to considering scenarios in which 
$\dot{R}<0$ in order for the formalism to apply, increasing time $t$ means 
decreasing $R(t)$ and {\it vice versa});

\item gravity ``cools'' if $f(R)$ grows faster than $R^2$ when 
$R$ increases (i.e., $t$ decreases);

\item pure $R^2$ gravity (which is in some sense pathological 
\cite{Pechlaner:1966dnt,Buchdahl70,Bicknell,Faraoni:2011pm}), is the 
threshold between these two behaviors, corresponding to the quadratic 
potential $V(\phi)=\alpha \phi^2/2$, which disappears from the field 
equation~(\ref{fe2}) for $\phi$. Then the electrovacuum  evolution 
equation for ${\cal K}{\cal T}$ reduces to 
\be
\frac{d \left( {\cal K}{\cal T}\right)}{d\tau} = {\cal K}{\cal T} 
 \left( 8\pi{\cal K}{\cal T} -\Theta \right)  \,, 
\ee  
which is the situation discussed in detail in Ref.~\cite{Faraoni:2025alq}. 

\end{itemize}

In the context of braneworld models, similar conclusions have been reached 
for Starobinsky $f(R)=R+\alpha R^2$ gravity on the brane 
\cite{Bhattacharyya:2025tgp}.

A comment is in order about the qualitative time evolution of $R(t)$, 
which coincides with that of $\phi(t)=f'(R(t))$ since $f'>0$.  In order to 
apply the thermal formalism we need decreasing $R(t)$. Since $R(t)$ is 
continuous and monotonically decreasing, it either goes to a horizontal 
asymptote $R_0$ as $t\to +\infty$ (where $R_0 \geq 0$ because the 
effective cosmological constant cannot be negative for the 
solutions~(\ref{y-solutions}), as already seen), or else $R(t)$ hits the 
$t$-axis from above with a non-zero derivative at a finite time $t_0$. 
Solutions with $R(t) \to +\infty$ as $t\to +\infty$ or to a finite time 
$t_0$ cannot be considered in the thermal formalism.

\section{FLRW universes in polynomial and power-law $f(R)$ gravity} 
\label{sec:4} 
\setcounter{equation}{0}

In this section, we consider polynomial functions 

\be 
f(R)=\sum_{n=1}^m  a_n R^n  \label{polynomial}
\ee 
(e.g., 
\cite{Barrow:1983rx,Cotsakis:1993zz,CotsakisFlessas1995,Cotsakis:2020kdl, 
Cotsakis:2013bza,Xavier:2020ulw}). $m=1$ corresponds to GR 
(possibly with a cosmological constant if we add a 
term $a_0\neq 0$ corresponding to $m=0$).  In general, 
there is no point in adding a cosmological constant to these theories 
since one of the main motivations for $f(R)$ gravity is to explain away 
dark energy and the fine-tuned cosmological constant. The recent DESI 
results point against a cosmological constant anyway 
\cite{DESI:2024aax,DESI:2024uvr,DESI:2024lzq}.

$m=2$ corresponds to quadratic quantum corrections to the Hilbert-Einstein 
action and to Starobinsky inflation $f(R)= R+\alpha R^2$ 
\cite{Starobinsky}. In the limit $ R\to 0$, the lowest power of 
$R$ in the sum~(\ref{polynomial}), corresponding to GR, dominates and one 
expects that, if the universe expands indefinitely with $R\to 0$ as it 
happens in classical GR solutions with $k=0, -1$, gravity converges to GR. 
Another possibility is that $R\to R_0 =$~const. from above, if de Sitter 
space is a late-time attractor in phase space.  (de Sitter space is a 
late-time attractor in the $\Lambda$CDM model and in many modified gravity 
cosmologies).  These are also ${\cal K}{\cal T}=0$ states in $f(R)$ gravity and 
one may still have convergence to GR in these situations---a case-by-case 
analysis is needed.

Conversely, if $R\to \infty$ approaching a singularity, 
the highest power of $R$ in the sum~(\ref{polynomial}) dominates. An 
infinite $R$ could be present as an initial  Big Bang singularity ($R(t) 
\to +\infty$ as $t\to 0^{+}$),\footnote{However, one can still have  
a Big Bang singularity without $R$ diverging, as it happens in the 
radiation era where 
$R$ vanishes identically.} as a 
final Big Crunch singularity ($R\to \infty$ as $t\to t_0^{-}$ with finite 
$t_0$), or as an asymptotic late-time 
singularity $R(t)\to -\infty$ as $t\to+\infty$. We do not discuss a  Big 
Rip singularity with $R(t)\to +\infty$ as $t \to t_0$, where $t_0$  
is finite,   because $R(t)$ must be decreasing in  order to apply the 
formalism.

There are many old results in the literature on scalar-tensor cosmology 
that are interesting for the thermal view and can be understood better and 
placed in a comprehensive framework using this new thermal formalism.

\begin{figure}
    \centering
    \includegraphics[width=0.85\linewidth]{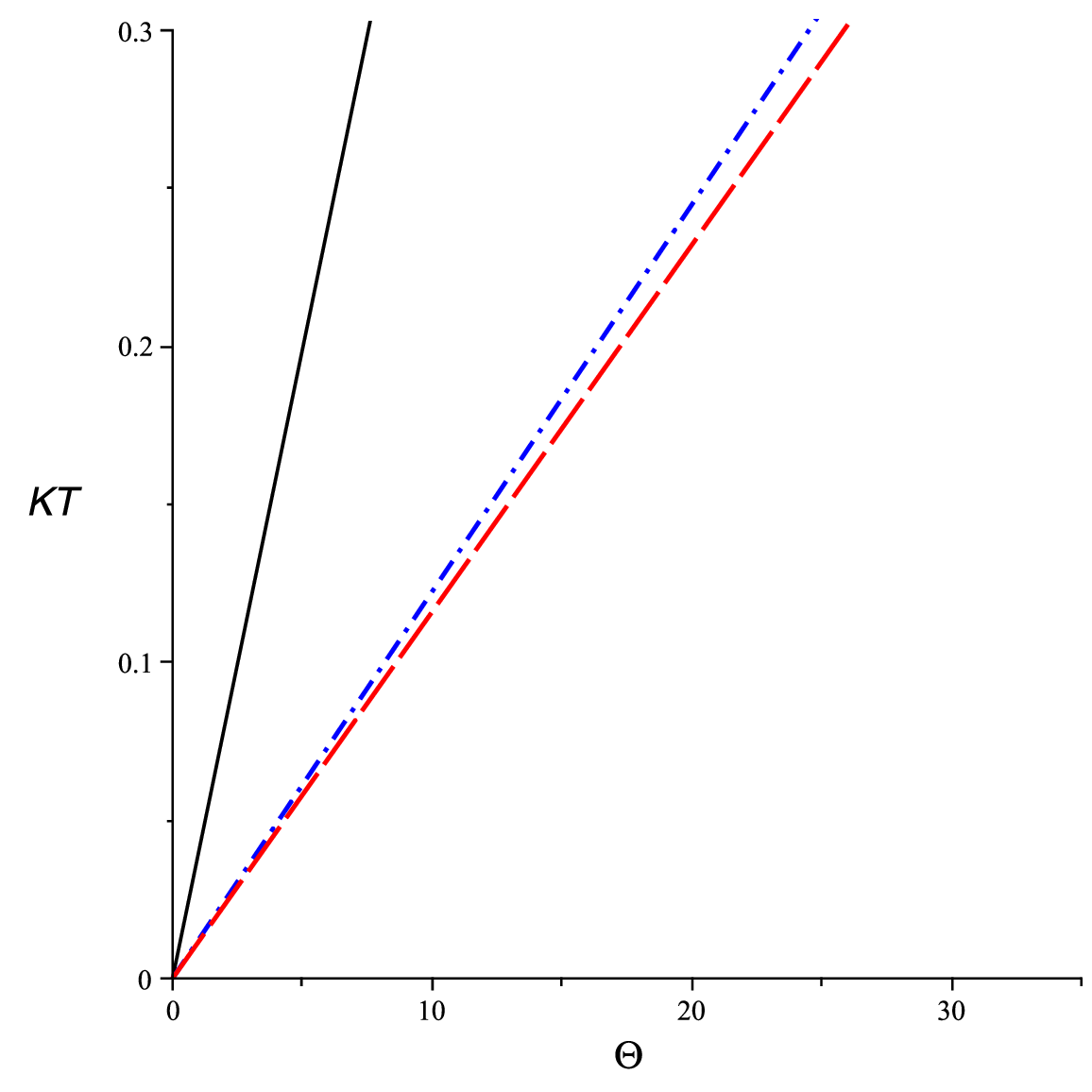}
\caption{In the region of  the $\left(\Theta, {\cal K}{\cal T} \right)$ 
plane between the half-line ${\cal K}{\cal T} = \Theta/(8\pi)$ (black, 
solid upper line) and the 
positive $\Theta$-axis, gravity ``cools'' and approaches GR. Outside of 
this region, gravity ``heats up'' and departs from GR. The red, dashed 
line corresponds to the solution~(\ref{eq:5.5})-(\ref{eq:5.6}) of 
vacuum $R^n$  gravity; the blue, 
dash-dotted line corresponds to the 
solution~(\ref{eq:5.15})-(\ref{eq:5.17}) 
of $R^n$ gravity with radiation. In both cases, $n=1.3$ and gravity moves 
towards the origin and GR.}
    \label{fig:plot-line}
\end{figure}

A body of work \cite{Ruzmaikina70,Cotsakis:1993zz, 
Barrow:1983rx, CotsakisFlessas1995,Cotsakis:2020kdl, Cotsakis:2013bza} 
(partially 
summarized in 
\cite{CotsakisFlessas1995})  studied FLRW 
cosmology in $f(R)$ gravity when 
the function $f(R)$ is polynomial.  With the general polynomial form of 
$f(R)$, the potential $V(\phi)$ for 
the scalar degree of freedom $\phi$ is complicated and does not resemble 
scalar field potentials motivated by high energy physics. However, if 
$f(R)$ 
consists of a single power-law $f(R)=R^n$ (or if the highest power in the 
sum dominates for large $R$, or the lowest power dominates as $R\to0$), 
then we know that the potential 
$V(\phi)=V_0 \phi^{\beta}$ contributes to ``cooling'' gravity if 
$\beta<2$, equivalent to $n>2$ \cite{Faraoni:2025alq}. {\em Vice-versa}, 
gravity is ``heated'' if $\beta >2$, or $ 1<n<2$. In the case of purely 
quadratic gravity $f(R)=R^2 $, which is a pathological theory in many 
respects \cite{Pechlaner:1966dnt, Buchdahl70, Bicknell, Faraoni:2011pm}, the 
potential $V(\phi)$ is quadratic and is well known not to affect the 
equation of motion of $\phi$, which it enters only in the combination 
$\phi dV/d\phi -2V$.  In this case, the evolution equation for 
${\cal K}{\cal T}$ reduces to 
\be
\frac{d\left( {\cal K}{\cal T}\right)}{d\tau}  = {\cal K}{\cal T} \left( 
8\pi {\cal K}{\cal T}-\Theta \right) +\frac{T^\mathrm{(m)} }{3f'(R)} 
\ee
and, for conformal matter with $T^\mathrm{(m)} =0$, it falls into the 
situation fully described in \cite{Faraoni:2025alq} (see 
Fig.~\ref{fig:plot-line}), which applies to general scalar-tensor gravity 
and not only to FLRW cosmology. If spacetime begins with initial 
conditions such that $\Theta >0$ and $8\pi {\cal K}{\cal T} <\Theta$, it 
approaches GR. If the initial conditions are such that $\Theta<0$, or 
$8\pi {\cal K}{\cal T}>\Theta>0$, gravity always runs away from GR due to 
the scalar degree of freedom dominating over the GR degrees of freedom 
\cite{Faraoni:2025alq}.

For the function $f(R)$ given by Eq.~(\ref{polynomial}), we have 
\begin{eqnarray}
\phi &=& f'(R) =\sum_{n=1}^m na_n R^{n-1} \,,\\
&&\nonumber\\
V(\phi) &=&  \sum_{n=2}^m (n-1) a_n R^n \,,\\
&&\nonumber\\
V'(\phi) &=& \frac{dV}{dR} \left( \frac{d\phi}{dR}\right)^{-1}   
=  \frac{\sum_{n=2}^m  n(n-1)a_n R^{n-1} }{ 
\sum_{n=2}^m n(n-1) a_n R^{n-2} } = R \,. \nonumber\\
&& 
\end{eqnarray}
Since 
\be
\phi V'-2V= 2f-\phi R = \sum_{n=1}^m (2-n)a_n R^n 
\ee
we have, for radiation or conformal matter with $T^\mathrm{(m)}=0$,
\begin{eqnarray}
\Box \phi &=& \frac{ \phi V'-2V}{3\phi} =\frac{1}{3} \left( 
\frac{2f}{\phi}-R \right) \nonumber\\
&&\nonumber\\
&=& \frac{ \sum_{n=1}^{m} (2-n)a_n R^n}{ 3\sum_{n=1}^m \, na_n R^{n-1}} 
\,.\label{urca}
\end{eqnarray}

The most important cases of polynomial $f(R)$ are power-law and 
Starobinsky gravity, which we examine next.

\section{Power-law $f(R)$}
\label{sec:5} 
\setcounter{equation}{0}

The case in which a single monomial  $R^n$ appears in the polynomial 
$f(R)$ has been 
studied extensively in attempts to explain the present acceleration of the universe 
without dark energy \cite{Errehymy:2024yey,Shubina:2021tgg,Ciftci:2017tjc, 
Deng:2014uta, Ganguly:2013taa, Schmidt:2012ms, Jaime:2012yi, 
DeBenedictis:2012qz,Gannouji:2011qz,Nojiri:2010wj,Nzioki:2010nj, 
Park:2010da,Leon:2010pu, Bisabr:2010sq, Capozziello:2009jg, 
Dunsby:2009zz,Faraoni:2009xb, Goheer:2009ss, 
Goheer:2008tn,Ananda:2008tx,Carloni:2008jy, AvilesCervantes:2008kno, 
Ananda:2008gs, Carloni:2007br,Goheer:2007wx,Ananda:2007xh, Martins:2007uf, 
Carloni:2007yv, Goheer:2007wu, Clifton:2007ih, Leach:2007ss, 
Capozziello:2007vd, Amendola:2006we,Carloni:2006mr, 
Clifton:2006kc,Carloni:2005ii, Capozziello:2006dp, Leach:2006br, 
Clifton:2006ug, Capozziello:2006ph, Capozziello:2006ph, Mendoza:2006hs, 
Sobouti:2006rd, Clifton:2005aj, Furey:2004rq,Pavlov:1997xf, 
Carloni:2004kp, CTT, Capozziello:2002rd, Ferraris:1992dx}. For $f(R)=R^n$, 
$f'>0 $ and $f''>0$ imply that $n > 1$ (if $R>0$), while 
\be
\phi = f'(R) = n R^{n-1} \,, \quad \quad 
R= \left( \frac{\phi}{n} \right)^{ \frac{1}{n-1} } \,,
\ee
and 
\begin{eqnarray}
V(\phi) &=&  Rf'(R)-f(R) \Bigg|_{\phi=f'(R)} = \left( n-1\right) R^n 
\nonumber\\
&&\nonumber\\
&=& \left( n-1 \right) \left( \frac{\phi}{n} \right)^{\frac{n}{n-1}} 
\end{eqnarray}
or
\be
V( \phi ) = V_0 \, \phi^\beta \,, \quad\quad \beta = \frac{n}{n-1} 
\ee
with 
\be
V_0 = \frac{ n-1}{ n^{ \frac{n}{n-1}} } \,.
\ee

If $\beta<2$, corresponding to $n>2$, the potential terms  ``cool'' 
gravity for $R>0$. If, 
instead, $\beta>2 $ (equivalent to $1<n<2$), they ``heat up'' gravity  
for $R>0$.

Let us look now at two examples using exact FLRW solutions of the field 
equations of Brans-Dicke gravity which are also solutions of $R^n$ 
gravity. The first exact solution \cite{Ciftci:2017tjc} corresponds to 
vacuum,  
$k=0$, a  power-law potential $V(\phi)=V_0 \phi^{\beta}$ and has 
power-law scale factor and scalar field
\begin{eqnarray}
a(t) &=& a_0 \, t^{ \frac{1+2\omega+\beta}{\left(2-\beta\right) \left( 
1-\beta \right)} } \equiv a_0 \, t^p \,, \label{eq:5.5}\\
&&\nonumber\\
\phi(t) &=& \phi_0 \, t^{ \frac{2}{1-\beta}} \,,\label{eq:5.6}
\end{eqnarray}
where $a_0$ and $\phi_0$ are positive constants. If we set $\omega=0$, 
this is also a solution 
of $R^n$ gravity with $\beta=\frac{n}{n-1}$ \cite{Ciftci:2017tjc}, 
provided that $n>1$ to keep $f''(R)>0$ (which also 
guarantees that $f'(R)>0$). This universe expands (i.e., 
$p\equiv \frac{(2n-1)(n-1)}{2-n} >0$) if $n<1/2$ or if $1<n<2$. The 
relevant range is, therefore, $1<n<2$. The scalar 
field derivative 
\be
\dot{\phi}=  -\, \frac{2\phi_0 \left( n-1 \right)}{ t^{2n-1}} 
\ee
is negative and so is $\dot{R}$, since
\be
H=\frac{p}{t} \,, \quad  R(t)= \frac{6p \left( 2p-1 \right)}{t^2} \,.
\ee
$R$ is positive when $p>1/2$, which corresponds to $5/4<n< 2 $, to which 
we restrict. We have 
\be
{\cal K}{\cal T} = \frac{n-1}{4\pi t} \to 0  \quad \mbox{as} \; t\to 
+\infty 
\ee
and 
\be
\Theta = 3H= \frac{3(2n-1)(n-1) }{ (2-n)t } \,,
\ee
so that 
\be
8\pi {\cal K}{\cal T} =\frac{2(2-n)}{3(2n-1)} \, \Theta \,.
\ee
We want to check whether the inequality $8\pi {\cal K}{\cal T}<\Theta $ is 
satisfied, which predicts the convergence to GR  ${\cal K}{\cal T} 
\to 0$. Indeed, for the coefficient of $\Theta$, $\frac{ 2(2-n) }{3(2n-1)} 
$,  
to be larger than unity it must be $n<7/8$, which is never satisfied in 
the selected range $5/4<n<2$, therefore, the thermal formalism predicts 
that this solution converges to GR at late times, which it does 
(Fig.~\ref{fig:plot-line}). 

Let us look now at another exact solution of Brans-Dicke gravity in the 
presence 
of a perfect  fluid with constant barotropic equation of state $P=\left( 
\gamma-1 \right)\rho$,  $V=V_0 \phi^{\beta}$, 
$k=0$, and \cite{Ciftci:2017tjc}
\begin{eqnarray}
a(t) &=& a_0 \, t^{ \frac{2\beta}{3\gamma \left(\beta -1\right)} }\,,\\
&&\nonumber\\
\phi(t) &=& \phi_0 \, t^{ \frac{2}{1-\beta } } \,,\\
&& \nonumber\\
\rho &=& \rho_0 \, t^{ \frac{2\beta}{1-\beta} } \,,
\end{eqnarray}
with constant $a_0,\phi_0, \rho_0$. Again, this is a solution of $R^n 
$ gravity if $\omega=0$ and $\beta=\frac{n}{n-1}$ with $n>1$, and we 
choose   a radiation fluid ($\gamma=4/3$) for which $T^\mathrm{(m)}=0$, 
obtaining
\begin{eqnarray}
a(t) &=& a_0 \, t^{ n/2 } \,, \label{eq:5.15}\\
&&\nonumber\\
\phi(t) &=& \phi_0 \, t^{ 2(1-n) } \,,\label{eq:5.16}\\
&& \nonumber\\
\rho &=& \rho_0 \, t^{-2n} \,.\label{eq:5.17}
\end{eqnarray}
For this solution, $H=n/(2t)$, $ R= \frac{3n(n-1)}{t^2} $
 and $\dot{R}<0$, therefore the thermal formalism 
is applicable. We have 
\be
8\pi {\cal K}{\cal T} = \frac{2(n-1)}{t}\to 0 \,, \quad \Theta= 
\frac{3n}{2t} \,, 
\ee
therefore,
\be
8\pi {\cal K}{\cal T} = \frac{ 4(n-1)}{3n} \Theta \label{urraurra} 
\ee
and the coefficient $\frac{ 4(n-1)}{3n}$ is less than unity for $1<n<4$. 
$\Theta $ starts infinite at $t=0$ and keeps decreasing, reaching zero as 
$t\to+\infty$. The point representing gravity in the $\left(\Theta, {\cal 
K}{\cal T}\right) $ plane moves along the straight line~(\ref{urraurra}) 
towards the origin (Fig.~\ref{fig:plot-line}).  

To understand this situation in detail, we look at the evolution equation 
for ${\cal K}{\cal T}$
\begin{eqnarray}
\frac{d\left( {\cal K}{\cal T}\right)}{dt} &=& 8\pi  
\left( {\cal K}{\cal T}\right)^2 -\Theta  {\cal K}{\cal T} 
+\frac{1}{24\pi} \left( V'-\frac{2V}{\phi}\right) \nonumber\\
&&\nonumber\\
&=& 8\pi  
\left( {\cal K}{\cal T}\right)^2 -\Theta  {\cal K}{\cal T} 
+\frac{V_0(2-n)  }{24\pi (n-1)}  \,  \phi^{ \frac{1}{n-1} } \,.\nonumber\\
&&
\end{eqnarray}
The left-hand side is $d \left( {\cal K}{\cal T}\right)/dt= 
-\frac{(n-1)}{4\pi t^2} $, while the right- hand side results from the 
competition of cooling and heating terms:
\begin{eqnarray}
&\, & {\cal K}{\cal T} \left( 8\pi {\cal K}{\cal T} -\Theta\right) 
+\frac{V_0}{24\pi} \left( \frac{2-n}{n-1}\right) n^{\frac{1}{n-1}}\, R 
\nonumber\\
&&\nonumber\\
&=& \frac{(n-1)(n-4)}{8\pi t^2} +  
\frac{(2-n)(n-1)}{8\pi t^2} = 
-\frac{(n-1)}{4\pi t^2}\nonumber\\
&& 
\end{eqnarray}
which, of course, equals the left-hand side and shows how the heating term 
(e.g., for $1<n<2$) 
\be
I \equiv 8\pi \left( {\cal K}{\cal T} \right)^2 + \frac{ (2-n)(n-1) }{8\pi t^2} =
\frac{(n-1)(3n-2)}{8\pi t^2}
\ee
 is smaller than the cooling term
\be
II \equiv \frac{ -3n(n-1) }{8\pi t^2 } \,.
\ee
In fact, the ratio
\be
\left| \frac{I}{II} \right|= \left| \frac{3n-2}{3n} \right|
\ee
is never larger than unity so the ``cooling'' term prevails for $ 1 <n<2$.

\subsection{Pure $R^2$-gravity}

The case $n=2$ of pure quadratic gravity $f(R)= R^2$ corresponds to 
$\beta=2$ and to a quadratic potential for the effective Brans-Dicke 
scalar. It is also the strong gravity regime of Starobinsky gravity 
discussed in the next section, described by $f(R)=R+\alpha R^2 \simeq 
\alpha R^2$ for large curvatures. In 
this special case the potential disappears from the equation of 
motion ~(\ref{fe2}) of $\phi$ and from Eq.~(\ref{evolution_general2}) and 
has no direct heating or cooling effect on gravity (cf. 
Eq.~(\ref{evolution_general2})). 

We have $\Box\phi=-\left( \ddot{\phi} + 3H \dot{\phi} \right)=0$, which 
admits the  first integral
\be
\dot{\phi} = \frac{C}{a^3}
\ee
well known in scalar-tensor cosmology. Since it must be $\dot{\phi}<0$ in 
order to apply the formalism, we restrict to values $C<0$ of the  
integration constant. On the other hand, $\phi=2R$ must be positive and 
the combined requirements $\phi>0$, $ \dot{\phi}<0$ give that $R>0$ must 
always decrease. Therefore, as $t \to +\infty$, $R$ approaches a 
horizontal asymptote $R(t)\to R_{\infty}^{+}$, or else $R$ vanishes at a 
finite time $t_0$.

If $R_{\infty}>0$, we have an asymptotic de Sitter equilibrium state 
analogous to GR, but de Sitter spaces in $R^2$ gravity are unstable with 
respect to both homogeneous and inhomogeneous perturbations 
(Appendix~\ref{appendix:A}), so this situation is physically irrelevant.

If $R_{\infty}=0$, we have an asymptotic Minkowski space corresponding to 
$\phi=2R \simeq 0$ and infinite effective gravitational coupling 
$G_\mathrm{eff} \simeq 1/\phi$, which is clearly away from GR. Indeed, it 
is well known that $R^2$ gravity does not admit a Newtonian limit around 
Minkowski space \cite{Pechlaner:1966dnt} (although it does admit one 
around de Sitter space \cite{Nguyen:2023whv}). Similarly, a Hamiltonian 
analysis reveals the absence of propagating degrees of freedom around 
Minkowski space, or any background with $R=0$, in the linear approximation 
(although the full theory propagates three degrees of freedom) 
\cite{Barker:2025gon}. Counting the propagating degrees of freedom is far 
from  trivial and depends on the background 
\cite{Hell:2023mph,Hell:2025wha,Hell:2025lbl}. 

In principle, the remaining possibility is the ``hard landing'' in which 
$R $ vanishes at a finite time $t_0$ with $\dot{R}<0$: in this case
\be
{\cal K}{\cal T}= \frac{ |\dot{R}|}{8\pi R} \to +\infty
\ee
and $R^2$ gravity runs infinitely far from GR.

\section{Starobinsky gravity}
\label{sec:6} 
\setcounter{equation}{0}

Quadratic terms in the Lagrangian density  are introduced  
by quantum corrections to GR, as in Starobinsky inflation 
\cite{Starobinsky},  
\be
f(R)=R+\alpha \, R^2 
\ee
with $\alpha>0$ to guarantee $f''>0$ and where $R> -1/(2\alpha)$ to 
keep $f'>0$. The trace equation~(\ref{metftrace}) 
reduces to \be
\Box R -\frac{R}{6\alpha} = \frac{4\pi}{3\alpha} \, T^\mathrm{(m)} \,.
\ee 
The other possible quadratic curvature  corrections 
$R_{ab}R^{ab}$ and 
$R_{abcd}R^{abcd}$ are contained implicitly in the Starobinsky action 
because, in four 
spacetime dimensions (to which we restrict), the Gauss-Bonnet term 
$R^2-4R_{ab}R^{ab}+R_{abcd}R^{abcd}$ is a topological invariant,
\be
\int d^4 x \, \sqrt{-g}  \left( 
R^2-4R_{ab}R^{ab}+R_{abcd}R^{abcd}\right)=\mbox{const.} \label{GaussBonnet}
\ee
and, additionally, because in FLRW spaces 
\be
\int d^4 x \, \sqrt{-g}  \left( R^2-3R_{ab}R^{ab} \right)=\mbox{const.} 
\label{DeWitt}
\ee
(e.g., \cite{DeWitt65}). Using Eqs.~(\ref{GaussBonnet}) 
and~(\ref{DeWitt}), one can trade 
$R_{abcd}R^{abcd}$ and $R_{ab}R^{ab}$ for terms in $R^2$ in the 
quadratic gravity action 
\be
S=
\int d^4x \, \sqrt{-g} \left( R+ \alpha R^2 +\beta R_{ab}R^{ab} +\gamma 
R_{abcd}R^{abcd} \right) \,.
\ee 
Quadratic gravity has been studied more extensively than 
other $f(R)$ theories. The Brans-Dicke theory equivalent to $f(R)=R+\alpha 
\, R^2 $ has $\omega=0$,
\be
\phi = 1+2\alpha R \,, \quad \quad R=\frac{1}{2\alpha} \left( \phi-1 
\right) \,,
\ee
and
\be
V(\phi) = \alpha R^2 = \frac{1}{4\alpha} \left( \phi-1\right)^2 \,. 
\label{StarobPotential}
\ee

There are no de Sitter spaces in Starobinsky gravity 
(Appendix~\ref{appendix:A}), hence the states ${\cal K}{\cal T}=$~const. 
can only be Minkowski spaces ($k=0$), Anti-de Sitter spaces ($k=-1$), or 
static Einstein universes ($k=+1$), if they exist, and they are physically 
relevant only if they are stable.\footnote{Thermal states of equilibrium 
with ${\cal K}{\cal T}=$~const.$>0$ are, in principle, possible in 
scalar-tensor gravity but, thus far, they have been shown to be unstable 
or unphysical \cite{Faraoni:2022doe, Giardino:2023qlu,Faraoni:2022fxo, 
Faraoni:2022jyd}).}

\begin{figure}
    \centering
    \includegraphics[width=0.85\linewidth]{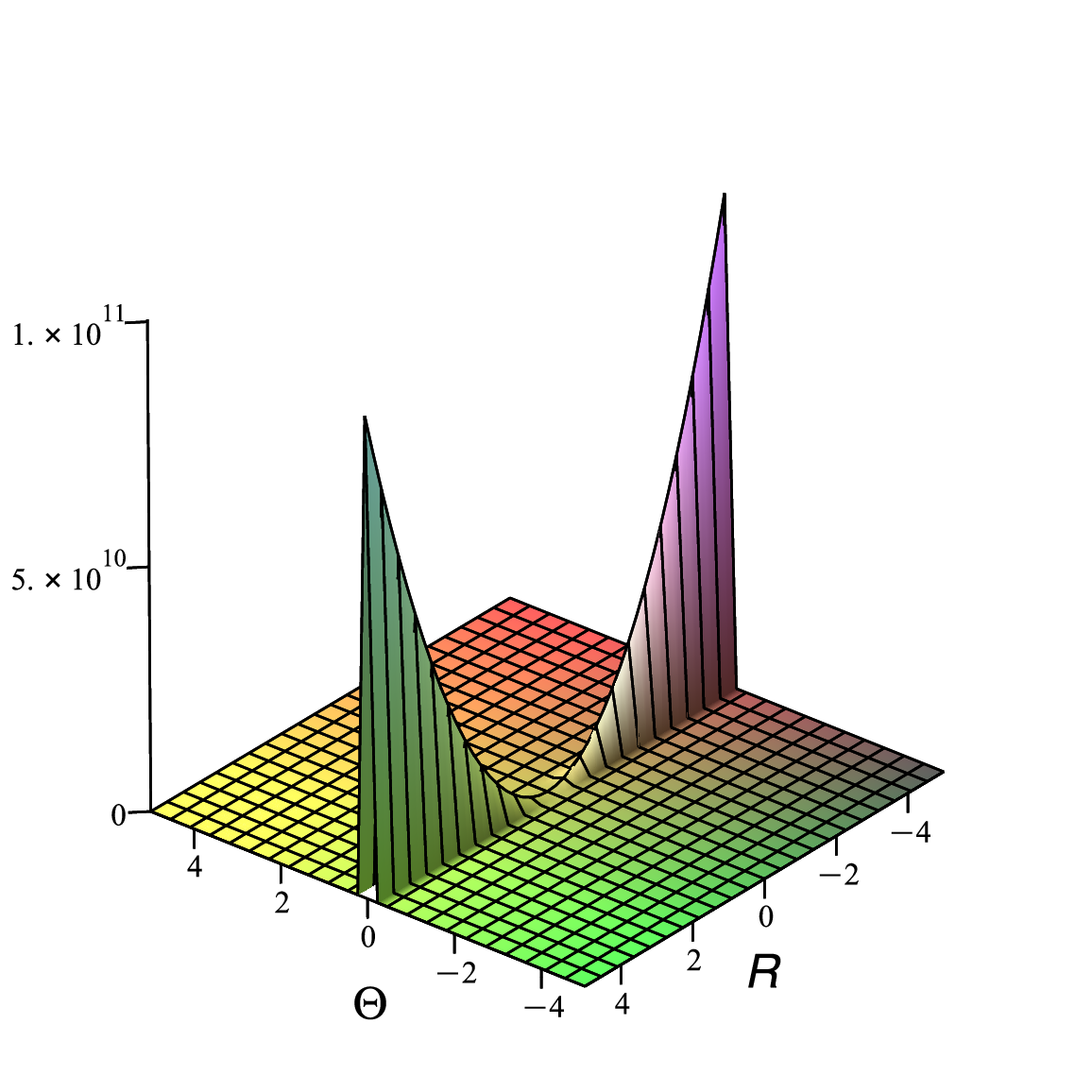}
    \caption{The orbits of the solutions of Starobinsky gravity are 
forced to lie on the surface ${\cal K}{\cal T} \left( \Theta,R 
\right)$ illustrated here in the 3D phase space $\left( \Theta, R, 
{\cal K}{\cal T} \right)$ (for illustration, we use $\alpha=10^{-3}$ and 
arbitrary units).}
    \label{fig:Starobinsky1}
\end{figure}

Assuming that the scalar degree of freedom dominates the cosmic dynamics, 
matter can be neglected and one can restrict to the vacuum field 
equations. The following discussion does not change if one considers a 
radiation fluid or other form of conformal matter with $T^\mathrm{(m)}=0$ 
(as done in \cite{CotsakisFlessas1995}). Then, the equation for ${\cal 
K}{\cal T}$ becomes
\be
\frac{ d\left( {\cal K}{\cal T}\right)}{d\tau} = {\cal K}{\cal T}\left( 8\pi {\cal 
K}{\cal T}-\Theta\right) + \frac{ \left( 1-1/\phi \right) }{48\pi\alpha} 
\,,
\ee
or
\be
\frac{ d\left( {\cal K}{\cal T}\right)}{d\tau} = {\cal K}{\cal T}\left( 
8\pi {\cal K}{\cal T}-\Theta\right) + \frac{ R}{24\pi \left( 1+2 \alpha R 
\right)} \,, \label{KTevolution-withR}  
\ee
while
\be
{\cal K}{\cal T}\left( \Theta, R \right)  = \frac{ \left| \alpha R^2 
-2\left(1+2\alpha  R\right)\Theta^2/3\right|}{16\pi \left( 1+2\alpha 
R\right)|\Theta| } \,.\label{urca2}
\ee
This surface in the $\left( \Theta, R, {\cal K}{\cal T} \right)$ space  
is plotted in Fig.~\ref{fig:Starobinsky1}. 

Since the scalar field potential~(\ref{StarobPotential}) has an 
absolute minimum at $\phi=1$, at late times the dynamics is attracted to 
this stable fixed point, which corresponds to $R\to 0$. In other words, as 
time progresses, the universe expands, $R$ decreases (making it possible 
to apply the thermal formalism), vanishing asymptotically as $t\to 
+\infty$. In this limit, the quadratic term in $f(R)=R+\alpha R^2$ becomes 
irrelevant and the theory reduces to GR (without cosmological constant)  
with $f(R) \simeq R$. In the 
thermal point of view, ${\cal K}{\cal T}\to 0^{+}$.

Let us discuss the late-time limit $R\to 0$ and the strong gravity regime 
$R\to \infty$.

\subsection{Limit $R\to0$}

If $R\to 0$, the scalar field degree of freedom freezes, i.e., $\phi 
=1+2\alpha R \to 1 $ and the potential disappears  from the evolution 
equation of ${\cal K}{\cal T}$: 
\be
\frac{ d\left( {\cal K}{\cal T}\right)}{d\tau} \simeq {\cal K}{\cal  
T}\left( 8\pi {\cal 
K}{\cal T}-\Theta\right) \,.
\ee
This  situation is described in  \cite{Faraoni:2025alq} and in 
Fig.~\ref{fig:plot-line}) for general  scalar-tensor gravity. Since 
$\dot{R}<0$, the limit $R\to 0$ implies that the axis $R=0$ is a 
horizontal asymptote for the function $R(t)$, with $\ddot{R}>0$. 

At small curvatures $R\to0$, the linear (GR) term dominates over the 
quadratic one in $f(R)=R+\alpha R^2$,
\be
 {\cal K}{\cal T} =  \frac{ |\dot{\phi}|}{8\pi \phi} 
=\frac{\alpha |\dot{R}|}{4\pi \left( 1+2\alpha R\right)} 
\simeq \frac{\alpha |\dot{R}|}{4\pi} \to 0 \,,
\ee
and GR is approached as $R(t)$ approaches its horizontal asymptote $R=0$. 
At this stage, if $R\to 0$ and assuming that it 
stays small,  Ref.~\cite{Faraoni:2025alq} and 
Fig.~\ref{fig:plot-line} tell us  that initial conditions such that the 
dynamics  begins 
below the half-line $8\pi 
{\cal K}{\cal T}=\Theta >0$ guarantee that gravity converges to GR while, 
if the 
universe begins above this line, or contracts with $\Theta<0$, it 
should run  away from GR. So why does Starobinsky gravity always 
converge to GR as the universe expands?  The key is that, as $R\to 0$, 
Eq.~(\ref{urca2}) yields 
\be
{\cal K}{\cal T} \left( \Theta, R \right) \simeq \frac{ 
|\Theta|}{24\pi} \to 0 \quad \mbox{as} \;\: \Theta\to 0
\ee
hence $8\pi {\cal K}{\cal T} \simeq |\Theta|/3 < \Theta $
and, in the late-time limit of large expansion  $R\to 0$, the point 
representing Starobinsky gravity in the $\left( \Theta, {\cal K}{\cal T} 
\right) $ plane always lies {\it below} the critical half-line $8\pi {\cal 
K}{\cal T}=\Theta$, so gravity is bound to converge to GR.

\subsection{Strong gravity regime $R\to\infty$}

In the strong gravity regime $R\to \infty$, $\phi \simeq 
2\alpha R \to \infty$ and 
\be 
\frac{ d\left( {\cal K}{\cal T}\right)}{d\tau} \simeq {\cal K}{\cal  
T}\left( 8\pi {\cal 
KT}-\Theta\right) + \frac{1}{48\pi\alpha} \,.\label{boh1} 
\ee 
The positive constant term in the right-hand side of Eq.~(\ref{boh1}) 
is a constant heat source which contributes to ``heating'' gravity and, if 
it was alone, it would cause 
${\cal K}{\cal T}$ to grow linearly in time. It is obvious that for contracting 
universes ($\Theta<0$), the right-hand side is positive and $d({\cal 
K}{\cal T})/d\tau>0$, hence gravity departs from GR. However, this is not the 
only situation when this happens.  

The constant curvature states with ${\cal K}{\cal T}=0$ are the {\it only} states 
with ${\cal K}{\cal T}=$~const. In fact, setting $d\left( {\cal 
K}{\cal T}\right)/d\tau=0$ (where $d\left( {\cal K}{\cal T}\right)/d\tau \equiv \psi 
\left( \Theta, {\cal K}{\cal T}\right)$) yields the formal algebraic 
equation
\be
\psi\left( {\cal K}{\cal T}, \Theta \right) \equiv 
8\pi \left( {\cal K}{\cal T}\right)^2 -\Theta {\cal 
K}{\cal T}+ \frac{1}{48\pi\alpha}=0 \label{formal}
\ee
with formal roots\footnote{Equation~(\ref{Roots}) generalizes the equation 
$ 8\pi 
{\cal K}{\cal T}-\Theta=0 $ used  in the corresponding analysis of thermal 
equilibrium states ${\cal K}{\cal T}=$~const. when 
$\Box\phi=0$ \cite{Faraoni:2025alq}.}  
\be
( {\cal K}{\cal T})_{\pm} ( \Theta) = \frac{1}{16\pi} \left( \Theta \pm \sqrt{ 
\Theta^2 
-\frac{2}{3\alpha} } \, \right) \geq 0  \label{Roots}
\ee
(which, if real,  are non-negative if $\Theta>0$). They depend, of course, 
on the coefficient $\Theta$ appearing in Eq.~(\ref{formal}). Obviously,  
${\cal K}{\cal T}$ can only 
be constant if this coefficient $\Theta =3H$ is 
constant, which only allows for de Sitter or Anti-de Sitter spaces. 
Minkowski space  and the Einstein static universe, which both have  
$\Theta=0$,  are ruled out 
since they 
make the argument of the square root in the right-hand side 
of~(\ref{Roots}) negative, while there are no de Sitter spaces in this 
theory (Appendix~\ref{appendix:A}).

\begin{figure}
    \centering
    \includegraphics[width=0.85\linewidth]{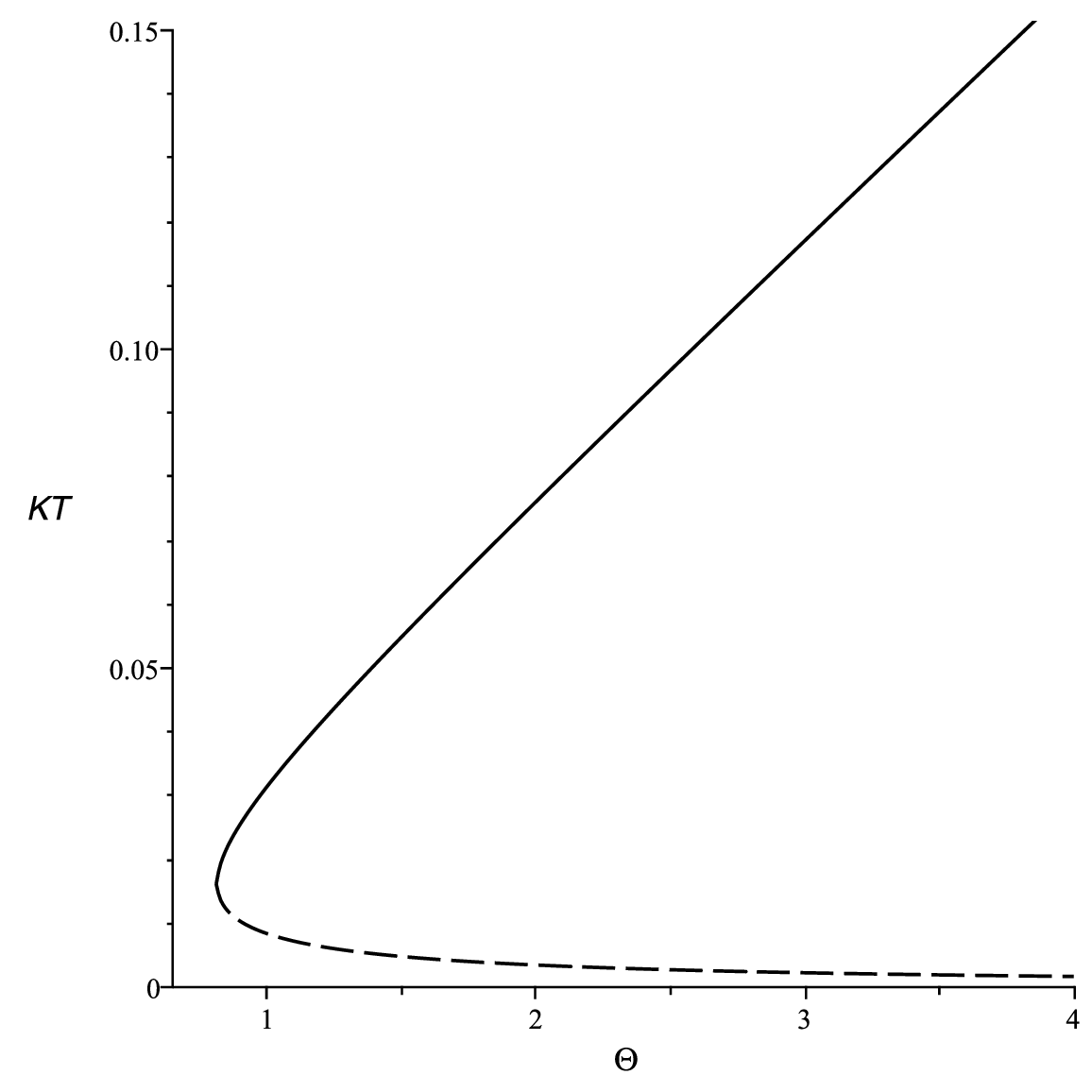}
    \caption{In the region of the $\left(\Theta, {\cal K}{\cal T} \right)$ 
plane between the graphs of the functions ${\cal 
K}{\cal T}_{+}(\Theta)$ (solid) and ${\cal 
K}{\cal T}_{-}(\Theta)$ (dashed), gravity 
``cools'' and approaches GR. Outside of this region, gravity ``heats 
up'' and departs from GR (here we set $\alpha=1$ for illustration).}
    \label{fig:plot}
\end{figure}

The function $ \psi \left( \Theta, {\cal K}{\cal T} \right) $ is negative for 
\be
\left( {\cal K}{\cal T} \right)_{-}(\Theta)  <  {\cal K}{\cal T} < 
\left( {\cal K} {\cal T} \right)_{+} ( \Theta) \,,
\ee
therefore, ${\cal K}{\cal T}$ decreases and gravity ``cools'' in the 
region of the $\left( \Theta, {\cal K}{\cal T} \right)$ plane comprised 
between the curves $ \left( {\cal K}{\cal T}\right)_{\pm}( \Theta)$ given 
by Eq.~(\ref{Roots}), while it ``heats up'' in the other two regions $ 
{\cal K}{\cal T}< ({\cal K}{\cal T})_{-} $ and $ {\cal K}{\cal T}> ({\cal 
K}{\cal T})_{+}$ of this plane (see Fig.~\ref{fig:plot}).

Instead of considering the  curves $ \left( {\cal 
K}{\cal T}\right)_{\pm}(\Theta)$, one could consider the inverse form
\be
\Theta \left( {\cal K}{\cal T}\right) = 8\pi {\cal K}{\cal 
T}  + \frac{1}{48\pi \alpha {\cal K}{\cal T}}  \,.
\ee
The curves $ \left( {\cal K}{\cal T}\right)_{\pm}(\Theta)$ themselves 
are not 
trajectories of 
the system. Next, one wonders whether  single 
points on these curves can be fixed points during 
time evolution. To answer, for  such a point to be a 
fixed point it must be $\Theta=$~const., while $d\left( {\cal 
K}{\cal T}\right) /dt =0$ by virtue of this point being on the curves 
$\left( {\cal K}{\cal T}\right)_{\pm}( \Theta)$. Then, the Ricci scalar
\begin{eqnarray}
R &=& 6\left( \dot{H} +2H^2 +\frac{k}{a^2} \right) 
= 6\left(  \frac{ \dot{\Theta}}{3} + \frac{2\Theta^2}{9} +\frac{k}{a^2} 
\right) \nonumber\\
&&\nonumber\\ 
&=& \frac{4\Theta^2}{3} +\frac{6k}{a^2}
\end{eqnarray}
is positive if $k=0, +1$ and a necessary (but not sufficient) condition 
for it to be negative is 
$k=-1$, i.e., that this point describes Anti-de Sitter space. However, 
this statement would contradict the regime under discussion because 
$\phi = 1+2\alpha R \simeq 2\alpha R$ when $R\to \infty$. Since $R<0$ for 
Anti-de Sitter space, the basic requirement that $\phi>0$ is violated. 
Therefore, Anti-de Sitter spaces on the curves ${\cal K}{\cal T}_{\pm}(\Theta) $ 
cannot be fixed 
points of the evolution and can, in principle, be crossed as time goes by. 

In quadratic gravity there are spatially flat, radiation-dominated 
($T^\mathrm{(m)}=0$) FLRW universes that avoid the initial Big Bang 
singularity and diverge in the future, $R(t) \to \infty $ as $t\to 
+\infty$ 
\cite{Ruzmaikina70,NariaiTomita1971,Giesswein:1974fog,Barrow:1983rx}. 
Clearly, as $R\to \infty$ it is $f(R)=R+\alpha R^2\simeq \alpha R^2$ and 
Starobinsky gravity departs drastically from GR (but the perturbative 
approach leading to Starobinsky gravity will break down at some point in 
this limit). These solutions are found to be unstable and, therefore, 
unphysical \cite{Barrow:1983rx}. In any case, they do not begin with 
$R=0$, but with a finite $R$. For these solutions, if $R(0)$ is larger 
than a critical value $R_*$, the solutions diverge as $t\to+\infty$, that 
is they diverge from GR.  If they begin with $R(0)<R_*$, then $R(t)\to 
0^{+}$ as $t\to+\infty$, approaching GR. The thermal view can be applied 
to these solutions with decreasing $R(t)$ and their behavior matches the 
results of the thermal view of scalar-tensor gravity 
\cite{Faraoni:2025alq} (applied to the specific subclass of Starobinsky 
gravity) and could have been predicted had this formalism been 
available in 1971. The thermal formalism cannot be applied to the 
solutions with $\dot{R}>0$.

Let us discuss now two other interesting classes of solutions of 
Starobinsky cosmology found numerically by Nariai and Tomita 
\cite{NariaiTomita1971}.  They studied the quadratic theory with action
\be 
S=\int d^4 x \sqrt{-g} \, \left[ R + \eta \left( 
R^2 +\alpha R_{ab}R^{ab} \right)\right] + S^\textrm{(m)} \,, 
\ee 
where matter is a radiation fluid, restricting themselves to $k=0$ FLRW 
solutions. Therefore, Eq.~(\ref{DeWitt}) applies and, {\em on FLRW 
solutions}, the theory can be recast as Starobinsky gravity 
$f(R)=R+\bar{\alpha} R^2 $, where $\bar{\alpha} = \eta \left( 1+\alpha/3 
\right)$. Nariai and Tomita looked for bouncing universes that avoid the 
Big Bang singularity \cite{NariaiTomita1971}.  Clearly, this is non-GR 
behavior because the radiation fluid always obeys the energy conditions 
and, in GR, a bounce with $\dot{H}>0$ requires the violation of the energy 
condition since $\dot{H}= -4\pi \left(\rho +P\right)$. This physics can 
be provided by the effective $\phi$-fluid which, being a fictitious fluid, 
can violate all of the energy conditions applicable to a real fluid. 
Nariai and Tomita integrated numerically the field equation for the scale 
factor $a(t)$ and found two types of solutions, which both begin by 
contracting ($\Theta=3H<0$) and reach an instant of time (which we will 
call $t=0$) at which $\dot{a}=0$ and $\Theta=3\dot{a}/a=0 $ 
\cite{NariaiTomita1971}.  At times $t>0$, one class of solutions expands 
forever, asymptoting to the GR solution $a(t) \simeq \sqrt{t}$ as 
$t\to+\infty$, which means that ${\cal K}{\cal T}\to 0^{+}$ while $R\to0$. The 
second class of solutions contracts again at $t>0$, ending in a Big Crunch 
with $\Theta\to-\infty$ and $R\to \infty$ at a finite time $t_0>0$ 
\cite{NariaiTomita1971}.  This means that $\ddot{a}$ changes from positive 
during the first contraction, to zero at $t=0$, to negative for $0<t<t_0$. 
This second type of solutions clearly departs from GR, but the thermal 
formalism is not applicable. In fact, for large $|R|$ it is $\phi \simeq 
2\alpha R$, which implies that it must be $R>0$. At the finite time 
singularity $R$ diverges so it must be  $R\to + \infty$. However, the 
Nariai-Tomita second class of solutions has $\dot{a}=0$ and $\ddot{a}=0$ 
at $t=0$, hence $R$ increases for times $0<t< t_0$, hence $ \dot{R}<0$ is 
not satisfied, $\dot{\phi}>0$, and the thermal formalism is not applicable. 

In the $\left( \Theta, {\cal K}{\cal T}\right)$ plane, these solutions of 
Starobinsky cosmology begin at negative $\Theta$ departing from GR and 
with upward-facing concavity, $\ddot{a}>0$.  In this region, $R=6\left( 
\frac{\ddot{a}}{a} + \frac{\dot{a}^2}{a^2} \right)$ is positive and 
Eq.~(\ref{KTevolution-withR}) becomes
\be
\frac{ d\left( {\cal K}{\cal T}\right)}{d\tau} = {\cal K}{\cal T}\left( 8\pi {\cal 
K}{\cal T} + |\Theta| \right) 
+ \frac{ |R|}{24\pi \left( 1+2 \alpha |R| \right)} \,; 
\ee 
the right-hand side is positive and ${\cal K}{\cal T}$ can only increase 
departing from GR. When the solution reaches the bounce at $\Theta=0$ 
(a strongly non-GR feature), the 
solutions  bifurcate: the first class enters a region where GR is 
approached and remains there, asymptotically approaching the GR solution 
$a(t)=a_0\sqrt{t}$. The second class of solutions goes back to the region 
$\Theta<0$, ending in a 
Big  Crunch as $\Theta $ goes towards $-\infty$, but the formalism 
loses meaning.  

Another result of quadratic gravity is that, if its spatially flat, 
radiation-dominated FLRW universes possess a Big Bang singularity, they 
approach the FLRW universes of GR as $t\to +\infty$ \cite{Barrow:1983rx}. 
This result is generalized by the Barrow-Ottewill theorem 
\cite{Barrow:1983rx} because Starobinsky gravity satisfies the hypothesis 
of the theorem $f(0)=0$ and $f'(0)\neq 0$.

\section{Conclusions} 
\label{sec:8} 
\setcounter{equation}{0}

The thermal view of scalar-tensor gravity is ultimately only an analogy 
between 
this class of alternative theories of gravity and heat dissipation in a 
viscous fluid. However, like all analogies, there are two sides to it and 
one can learn much about gravity from physical intuition of the other 
side. Physical insight has been obtained in first-generation and 
Horndeski 
gravities, and here we have applied the thermal analogy specifically to 
$f(R)$ gravity, which is very popular to explain the current acceleration 
of the universe, provides the first scenario of inflation ever proposed 
(by Starobinsky), which is very successful, and is motivated by quadratic 
corrections to GR.

The main value of the thermal view of scalar-tensor gravity consists of 
its unifying power. The analysis of the previous sections has framed known 
results, some of which are old and remained rather cryptic, in the thermal 
view of $f(R)$ gravity and cosmology. These results remained disconnected 
for decades, and now fall into a single, coherent picture. The fourth 
order field equations of $f(R)$ gravity are rather complicated, even for 
spatially homogeneous and isotropic FLRW universes, and there is a large 
variety of situations corresponding to different matter contents of the 
universe, forms of the function $f(R)$, and initial conditions. It is 
logical, therefore, that one cannot catch all these possible theories and 
scenarios in a few single sentences. We find significant, however, that 
the new thermal view of $f(R)$ gravity unifies many apparently fragmented, 
and often forgotten, results, as discussed in the previous sections.

The results concerning scalar-tensor and $f(R)$ universes that we discussed here 
can be obtained by studying the dynamics ruled by the master equations of these 
theories (i.e., Eqs.~\ref{metf}), (\ref{metftrace})) and, mathematically, do not 
require the thermal formalism (i.e., Eq.~(\ref{KTf(R)})) to be deduced. However, in 
many situations one would not think of investigating certain directions unless 
guided by the thermal view, and one often lacks a physical interpretation of these 
results without the unifying thermal framework.

Here we focussed on FLRW cosmologies and on expanding universes and we did 
not delve into the features of static or contracting universes, which are 
less interesting from the physical point of view, and will be presented  
elsewhere.  There are also several results in the literature on spatially 
anisotropic Bianchi  universes in $f(R)$ and in more general Horndeski 
cosmology, which seem relevant for the thermal view of scalar-tensor 
cosmology and will be analyzed separately.

\begin{acknowledgments}

This work is supported, in part, by the Natural Sciences \& Engineering 
Research Council of Canada (grant 2023-03234 to V.~F.).

\end{acknowledgments}

\begin{appendices} 
\section{Existence and stability of de Sitter spaces in 
vacuum $f(R)$ gravity} 
\label{appendix:A} 
\renewcommand{\theequation}{A.\arabic{equation}} \setcounter{equation}{0} 

The condition for the existence of de Sitter spaces with Hubble function  
$H_0=$~const. and Ricci scalar $R_0=12H_0^2$ in vacuum $f(R)$ gravity 
is derived by substituting the de Sitter metric into Eq.~(\ref{metf}), 
obtaining  \cite{Barrow:1983rx,Faraoni:2005vk}
\be
R_0 f'_0 = 2f_0 \,,\label{existence}
\ee
where $f_0\equiv f(R_0)$ and $f'_0 \equiv f'(R_0)$. If such a de Sitter 
space exists, the condition for its stability with respect to both 
homogeneous and inhomogeneus perturbations (to first order) is 
\cite{Faraoni:2005vk}
\be
\left( f'_0 \right)^2 - 2f_0 f_0'' \geq 0 \,. \label{stability}
\ee

For Starobinsky gravity with $f(R)=R+\alpha R^2$, the existence 
condition~(\ref{existence}) cannot be satisfied and there are no de Sitter 
solutions in this theory.

In $R^n$ gravity, the existence condition for de Sitter spaces can be 
satisfied only if $n=2$. 
In this case, however, the stability condition becomes $-R^2 \geq 0$, 
which cannot be satisfied except for the degenerate Minkowski solution 
with $H_0=0$ and $R_0=0$ (a trivial de Sitter space).


\end{appendices}


\begin{thebibliography}{99}
 


\bibitem{Stelle:1976gc} K.~S.~Stelle, ``Renormalization of Higher 
Derivative Quantum Gravity,'' Phys. Rev. D \textbf{16}, 953-969 (1977) 
doi:10.1103/PhysRevD.16.953

\bibitem{Stelle:1977ry} K.~S.~Stelle, ``Classical Gravity with Higher 
Derivatives,'' Gen. Rel. Grav. \textbf{9}, 353-371 (1978) 
doi:10.1007/BF00760427

\bibitem{Callan:1985ia} C.~G.~Callan, Jr., E.~J.~Martinec, M.~J.~Perry and 
D.~Friedan, ``Strings in Background Fields,'' Nucl. Phys. B \textbf{262}, 
593-609 (1985) doi:10.1016/0550-3213(85)90506-1

\bibitem{Fradkin:1985ys} E.~S.~Fradkin and A.~A.~Tseytlin, ``Quantum 
String Theory Effective Action,'' Nucl. Phys. B \textbf{261}, 1-27 (1985) 
[erratum: Nucl. Phys. B \textbf{269}, 745-745 (1986)] 
doi:10.1016/0550-3213(85)90559-0

\bibitem{Brans:1961sx} C.~Brans and R.~H.~Dicke, ``Mach's principle and a 
relativistic theory of gravitation'', Phys. Rev. \textbf{124}, 925-935 
(1961) doi:10.1103/PhysRev.124.925.

\bibitem{Bergmann:1968ve} P.~G.~Bergmann, ``Comments on the scalar tensor 
theory'', Int. J. Theor. Phys. \textbf{1}, 25-36 (1968)
 doi:10.1007/BF00668828.

\bibitem{Nordtvedt:1968qs} K.~Nordtvedt, ``Equivalence Principle for 
Massive Bodies. 2. Theory'', Phys. Rev. \textbf{169}, 1017-1025 (1968)
 doi:10.1103/PhysRev.169.1017.

\bibitem{Wagoner:1970vr} R.~V.~Wagoner, ``Scalar tensor theory and 
gravitational waves'', Phys. Rev. D \textbf{1}, 3209-3216 (1970)
 doi:10.1103/PhysRevD.1.3209.

\bibitem{Nordtvedt:1970uv} K.~Nordtvedt, Jr., ``PostNewtonian metric for a 
general class of scalar tensor gravitational theories and observational 
consequences'', Astrophys. J. \textbf{161}, 1059-1067 (1970)
 doi:10.1086/150607.

\bibitem{Starobinsky} A.~A.~Starobinsky, ``A New Type of Isotropic 
Cosmological Models Without Singularity,'' Phys. Lett. B \textbf{91}, 
99-102 (1980), doi:10.1016/0370-2693(80)90670-X

\bibitem{Amendola:2015ksp} L.~Amendola and S.~Tsujikawa, {\it Dark Energy: 
Theory and Observations}, Cambridge University Press, 2015, ISBN 
978-1-107-45398-2

\bibitem{Riess:2019qba} A.~G.~Riess, ``The Expansion of the Universe is 
Faster than Expected,'' Nature Rev. Phys. \textbf{2}, no.1, 10-12 (2019) 
doi:10.1038/s42254-019-0137-0 [arXiv:2001.03624 [astro-ph.CO]].

\bibitem{DiValentino:2021izs} E.~Di Valentino, O.~Mena, S.~Pan, 
L.~Visinelli, W.~Yang, A.~Melchiorri, D.~F.~Mota, A.~G.~Riess and J.~Silk, 
``In the realm of the Hubble tension\textemdash{}a review of solutions,'' 
Class. Quant. Grav. \textbf{38}, no.15, 153001 (2021) 
doi:10.1088/1361-6382/ac086d [arXiv:2103.01183 [astro-ph.CO]].

\bibitem{Sotiriou:2008rp}
T.~P.~Sotiriou and V.~Faraoni,
``f(R) Theories Of Gravity,''
Rev. Mod. Phys. \textbf{82}, 451-497 (2010)
doi:10.1103/RevModPhys.82.451
[arXiv:0805.1726 [gr-qc]].

\bibitem{DeFelice:2010aj}A.~De Felice and S.~Tsujikawa, ``$f(R)$ 
theories,'' Living Rev. Rel. \textbf{13}, 3 (2010) 
doi:10.12942/lrr-2010-3[arXiv:1002.4928 [gr-qc]].

\bibitem{Nojiri:2010wj} S.~Nojiri and S.~D.~Odintsov,``Unified cosmic 
history in modified gravity: from $F(R)$ theory to Lorentz non-invariant 
models,'' Phys. Rept. \textbf{505}, 59-144 (2011) 
doi:10.1016/j.physrep.2011.04.001[arXiv:1011.0544 [gr-qc]].

\bibitem{Horndeski:1974wa} G.~W.~Horndeski, ``Second-order scalar-tensor 
field equations in a four-dimensional space,'' Int. J. Theor. Phys. 
\textbf{10}, 363-384 (1974) doi:10.1007/BF01807638.

\bibitem{Faraoni:2018qdr} V.~Faraoni and J.~C\^ot\'e, ``Imperfect fluid 
description of modified gravities,'' Phys. Rev. D \textbf{98} no.~8, 
084019 (2018) doi:10.1103/PhysRevD.98.084019 [arXiv:1808.02427 [gr-qc]].

\bibitem{Faraoni:2021lfc} V.~Faraoni and A.~Giusti, ``Thermodynamics of 
scalar-tensor gravity,'' Phys. Rev. D \textbf{103}, no.12, L121501 (2021) 
doi:10.1103/PhysRevD.103.L121501 [arXiv:2103.05389 [gr-qc]].

\bibitem{Faraoni:2021jri} V.~Faraoni, A.~Giusti and A.~Mentrelli, ``New 
approach to the thermodynamics of scalar-tensor gravity,'' Phys. Rev. D 
\textbf{104}, no.12, 124031 (2021) doi:10.1103/PhysRevD.104.124031 
[arXiv:2110.02368 [gr-qc]].

\bibitem{Giusti:2021sku} A.~Giusti, S.~Zentarra, L.~Heisenberg and 
V.~Faraoni, ``First-order thermodynamics of Horndeski gravity,'' Phys. 
Rev. D \textbf{105}, no.12, 124011 (2022) doi:10.1103/PhysRevD.105.124011 
[arXiv:2108.10706 [gr-qc]].

\bibitem{Faraoni:2022gry} V.~Faraoni, S.~Giardino, A.~Giusti and 
R.~Vanderwee, ``Scalar field as a perfect fluid: thermodynamics of 
minimally coupled scalars and Einstein frame scalar-tensor gravity,'' Eur. 
Phys. J. C \textbf{83}, no.1, 24 (2023)
 doi:10.1140/epjc/s10052-023-11186-7 [arXiv:2208.04051 [gr-qc]].

\bibitem{Giardino:2022sdv} S.~Giardino, V.~Faraoni and A.~Giusti, 
``First-order thermodynamics of scalar-tensor cosmology,'' JCAP 
\textbf{04}, no.04, 053 (2022) doi:10.1088/1475-7516/2022/04/053 
[arXiv:2202.07393 [gr-qc]].

\bibitem{Gallerani:2024gdy} L.~Gallerani, M.~Miranda, A.~Giusti and 
A.~Mentrelli, ``Alternative formulations of the thermodynamics of 
scalar-tensor theories,'' Phys. Rev. D \textbf{110}, no.6, 064087 (2024) 
doi:10.1103/PhysRevD.110.064087 [arXiv:2405.20865 [gr-qc]].

\bibitem{Faraoni:2025alq}
V.~Faraoni and A.~Giusti,
``Thermal Origin of the Attractor-to-General-Relativity in Scalar-Tensor 
Gravity,''
Phys. Rev. Lett. \textbf{134}, no.21, 211406 (2025)
doi:10.1103/22w4-v2xn
[arXiv:2502.18272 [gr-qc]].

\bibitem{Faraoni:2025ufi}
V.~Faraoni,
``Black hole interiors in the thermal view of scalar-tensor gravity,''
Phys. Rev. D \textbf{112}, no.2, L021504 (2025)
doi:10.1103/mk89-hjkn
[arXiv:2505.08322 [gr-qc]].
 
\bibitem{Pereira:2025dmk}
D.~S.~Pereira and J.~P.~Mimoso,
``Eckart heat-flux applicability in $F(\Phi , X) R$ theories and the existence of 
temperature gradients,''
[arXiv:2512.20553 [gr-qc]].

\bibitem{Damour:1992kf}
T.~Damour and K.~Nordtvedt,
``General relativity as a cosmological attractor of tensor scalar 
theories,''
Phys. Rev. Lett. \textbf{70}, 2217-2219 (1993) 
doi:10.1103/PhysRevLett.70.2217

\bibitem{Damour:1993id}
T.~Damour and K.~Nordtvedt,
``Tensor-scalar cosmological models and their relaxation toward general 
relativity,''
Phys. Rev. D \textbf{48}, 3436-3450 (1993) doi:10.1103/PhysRevD.48.3436

\bibitem{Serna:2002fj}
A.~Serna, J.~M.~Alimi and A.~Navarro,
``Convergence of scalar tensor theories toward general relativity and 
primordial nucleosynthesis,''
Class. Quant. Grav. \textbf{19}, 857-874 (2002)
doi:10.1088/0264-9381/19/5/302 [arXiv:gr-qc/0201049 [gr-qc]].

\bibitem{Wald:1984rg} R.~M.~Wald, {\it General Relativity}, Chicago 
University Press, 1984, doi:10.7208/chicago/9780226870373.001.0001

\bibitem{Eckart:1940te} C.~Eckart, ``The thermodynamics of irreversible 
processes. 3.~Relativistic theory of the simple fluid,'' Phys. Rev. 
\textbf{58}, 919-924 (1940), doi:10.1103/PhysRev.58.919

\bibitem{Maartens:1996vi} R.~Maartens, ``Causal thermodynamics in 
relativity,'' [arXiv:astro-ph/9609119 [astro-ph]].

\bibitem{Andersson:2006nr} N.~Andersson and G.~L.~Comer, ``Relativistic 
fluid dynamics: Physics for many different scales,'' Living Rev. Rel. 
\textbf{10}, 1 (2007), doi:10.12942/lrr-2007-1 [arXiv:gr-qc/0605010 
[gr-qc]].

\bibitem{Faraoni:2023hwu}
V.~Faraoni and J.~Houle,
``More on the first-order thermodynamics of scalar-tensor and Horndeski 
gravity,''
Eur. Phys. J. C \textbf{83}, no.6, 521 (2023)
doi:10.1140/epjc/s10052-023-11712-7
[arXiv:2302.01442 [gr-qc]].

\bibitem{Faraoni:2025dex}
V.~Faraoni and N.~Veilleux,
``The thermal view of singularity-free scalar-tensor spacetimes,''
[arXiv:2511.04941 [gr-qc]].

\bibitem{Cognola:2005de}
G.~Cognola, E.~Elizalde, S.~Nojiri, S.~D.~Odintsov and S.~Zerbini,
``One-loop f(R) gravity in de Sitter universe,''
JCAP \textbf{02}, 010 (2005)
doi:10.1088/1475-7516/2005/02/010
[arXiv:hep-th/0501096 [hep-th]].

\bibitem{Nojiri:2006gh}
S.~Nojiri and S.~D.~Odintsov,
``Modified f(R) gravity consistent with realistic cosmology: From matter 
dominated epoch to dark energy universe,''
Phys. Rev. D \textbf{74}, 086005 (2006)
doi:10.1103/PhysRevD.74.086005
[arXiv:hep-th/0608008 [hep-th]].

\bibitem{Capozziello:2006dj}
S.~Capozziello, S.~Nojiri, S.~D.~Odintsov and A.~Troisi,
``Cosmological viability of f(R)-gravity as an ideal fluid and its 
compatibility with a matter dominated phase,''
Phys. Lett. B \textbf{639}, 135-143 (2006)
doi:10.1016/j.physletb.2006.06.034
[arXiv:astro-ph/0604431 [astro-ph]].

\bibitem{Cognola:2008zp}
G.~Cognola, E.~Elizalde, S.~D.~Odintsov, P.~Tretyakov and S.~Zerbini,
``Initial and final de Sitter universes from modified f(R) gravity,''
Phys. Rev. D \textbf{79}, 044001 (2009)
doi:10.1103/PhysRevD.79.044001
[arXiv:0810.4989 [gr-qc]].

\bibitem{Nojiri:2009kx}
S.~Nojiri, S.~D.~Odintsov and D.~Saez-Gomez,
``Cosmological reconstruction of realistic modified F(R) gravities,''
Phys. Lett. B \textbf{681}, 74-80 (2009)
doi:10.1016/j.physletb.2009.09.045
[arXiv:0908.1269 [hep-th]].

\bibitem{Dolgov:2003px}
A.~D.~Dolgov and M.~Kawasaki,
``Can modified gravity explain accelerated cosmic expansion?,''
Phys. Lett. B \textbf{573}, 1-4 (2003)
doi:10.1016/j.physletb.2003.08.039
[arXiv:astro-ph/0307285 [astro-ph]].

\bibitem{Faraoni:2006sy}
V.~Faraoni,
``Matter instability in modified gravity,''
Phys. Rev. D \textbf{74}, 104017 (2006)
doi:10.1103/PhysRevD.74.104017
[arXiv:astro-ph/0610734 [astro-ph]].

\bibitem{Giardino:2023qlu} S.~Giardino, A.~Giusti and V.~Faraoni, 
``Thermal stability of stealth and de Sitter spacetimes in scalar-tensor 
gravity,'' Eur. Phys. J. C \textbf{83}, no.7, 621 (2023) 
doi:10.1140/epjc/s10052-023-11697-3 [arXiv:2302.08550 [gr-qc]].

\bibitem{Faraoni:2005vc}
V.~Faraoni,
``Phase space geometry in scalar-tensor cosmology,''
Annals Phys. \textbf{317}, 366-382 (2005)
doi:10.1016/j.aop.2004.11.009
[arXiv:gr-qc/0502015 [gr-qc]].

\bibitem{deSouza:2007zpn}
J.~C.~C.~de Souza and V.~Faraoni,
``The Phase space view of f(R) gravity,''
Class. Quant. Grav. \textbf{24}, 3637-3648 (2007)
doi:10.1088/0264-9381/24/14/006
[arXiv:0706.1223 [gr-qc]].

\bibitem{Barrow:1983rx}
J.~D.~Barrow and A.~C.~Ottewill,
``The Stability of General Relativistic Cosmological Theory,''
J. Phys. A \textbf{16}, 2757 (1983)
doi:10.1088/0305-4470/16/12/022

\bibitem{CotsakisFlessas1995} Cotsakis, S. and Flessas, G., 1995. 
Past-instability conjecture and cosmological attractors in generalized 
isotropic universes. Physical Review D, 51(8), p.416

\bibitem{Hervik:2017sdr}
S.~Hervik, V.~Pravda and A.~Pravdov{\'a},
``Universal spacetimes in four dimensions,''
JHEP \textbf{10}, 028 (2017)
doi:10.1007/JHEP10(2017)028
[arXiv:1707.00264 [gr-qc]]

\bibitem{Hervik:2013cla}
S.~Hervik, V.~Pravda and A.~Pravdova,
``Type III and N universal spacetimes,''
Class. Quant. Grav. \textbf{31}, no.21, 215005 (2014)
doi:10.1088/0264-9381/31/21/215005
[arXiv:1311.0234 [gr-qc]].

\bibitem{Faraoni:2021opj}
V.~Faraoni, S.~Jose and S.~Dussault,
``Multi-fluid cosmology in Einstein gravity: analytical solutions,''
Gen. Rel. Grav. \textbf{53}, no.12, 109 (2021)
doi:10.1007/s10714-021-02879-z
[arXiv:2107.12488 [gr-qc]].

\bibitem{Ruban:1972bg} V.~A.~Ruban and A.~M.~Finkelstein, 
``Generalization of the taub-kazner cosmological metric in the 
scalar-tensor gravitation theory,'' 
Lett. Nuovo Cim. \textbf{5S2}, 289-293 (1972) 
doi:10.1007/BF02752628

\bibitem{Ruzmaikina70} T. Ruzmaikina and A.~A. Ruzmaikin, ``Quadratic 
corrections to the Lagrangian density of the gravitational field and the 
singularity'', Sov. Phys. JETP {\bf 30}, 372 (1970)

\bibitem{Cotsakis:1993zz}
S.~Cotsakis and G.~Flessas,
``Stability of FRW cosmology in higher order gravity,''
Phys. Rev. D \textbf{48}, 3577-3584 (1993)
doi:10.1103/PhysRevD.48.3577

\bibitem{Hu:2007nk}
W.~Hu and I.~Sawicki,
``Models of f(R) Cosmic Acceleration that Evade Solar-System Tests,''
Phys. Rev. D \textbf{76} (2007), 064004
doi:10.1103/PhysRevD.76.064004
[arXiv:0705.1158 [astro-ph]].

\bibitem{Starobinsky:2007hu}
A.~A.~Starobinsky,
``Disappearing cosmological constant in f(R) gravity,''
JETP Lett. \textbf{86} (2007), 157-163
doi:10.1134/S0021364007150027
[arXiv:0706.2041 [astro-ph]].

\bibitem{Miranda:2009rs}
V.~Miranda, S.~E.~Joras, I.~Waga and M.~Quartin,
``Viable Singularity-Free f(R) Gravity Without a Cosmological Constant,''
Phys. Rev. Lett. \textbf{102} (2009), 221101
doi:10.1103/PhysRevLett.102.221101
[arXiv:0905.1941 [astro-ph.CO]].

\bibitem{Pechlaner:1966dnt}
E.~Pechlaner and R.~Sexl,
``On quadratic lagrangians in General Relativity,''
Commun. Math. Phys. \textbf{2}, no.1, 165-175 (1966)
doi:10.1007/BF01773351

\bibitem{Buchdahl70} H. A. Buchdahl, ``Non-linear Lagrangians and 
cosmological theory'', Mon.  Not. Royal Astron. Soc.  {\textbf 150}, 1--8 
(1970)

\bibitem{Bicknell} G. V. Bicknell, ``Non-viability of gravitational theory 
based on a quadratic lagrangian'', J. Phys.~A: Mathematical, Nuclear and 
General {\textbf 7}, 1061 (1974)

\bibitem{Faraoni:2011pm}
V.~Faraoni, ``$R^n$ gravity and the chameleon,''
Phys. Rev. D \textbf{83}, 124044 (2011)
doi:10.1103/PhysRevD.83.124044
[arXiv:1106.0328 [gr-qc]].

\bibitem{Bhattacharyya:2025tgp}
S.~Bhattacharyya and S.~SenGupta,
``Thermal description of braneworld effective theories,''
[arXiv:2508.14228 [hep-th]].

\bibitem{Cotsakis:2020kdl}
S.~Cotsakis and D.~Trachilis,
``The radiation instability in modified gravity,''
Int. J. Mod. Phys. A \textbf{36}, no.08n09, 2150060 (2021)
doi:10.1142/S0217751X21500603
[arXiv:2012.05850 [gr-qc]].

\bibitem{Cotsakis:2013bza}
S.~Cotsakis, D.~Trachilis and A.~Tsokaros,
``Generic regular universes in higher order gravity theories,''
doi:10.1142/9789814623995\_0301
[arXiv:1302.6674 [gr-qc]].

\bibitem{Xavier:2020ulw}
S.~Xavier, J.~Mathew and S.~Shankaranarayanan,
``Infinitely degenerate exact Ricci-flat solutions in f(R) gravity,''
Class. Quant. Grav. \textbf{37}, no.22, 225006 (2020)
doi:10.1088/1361-6382/abbd0f
[arXiv:2003.05139 [gr-qc]].

\bibitem{DESI:2024aax}
A.~G.~Adame \textit{et al.} [DESI],
``DESI 2024 II: sample definitions, characteristics, and two-point 
clustering statistics,''
JCAP \textbf{07}, 017 (2025)
doi:10.1088/1475-7516/2025/07/017
[arXiv:2411.12020 [astro-ph.CO]].

\bibitem{DESI:2024uvr}
A.~G.~Adame \textit{et al.} [DESI],
``DESI 2024 III: baryon acoustic oscillations from galaxies and 
quasars,''
JCAP \textbf{04}, 012 (2025)
doi:10.1088/1475-7516/2025/04/012
[arXiv:2404.03000 [astro-ph.CO]].

\bibitem{DESI:2024lzq}
A.~G.~Adame \textit{et al.} [DESI],
``DESI 2024 IV: Baryon Acoustic Oscillations from the Lyman alpha 
forest,''
JCAP \textbf{01}, 124 (2025)
doi:10.1088/1475-7516/2025/01/124
[arXiv:2404.03001 [astro-ph.CO]].

\bibitem{Errehymy:2024yey}
A.~Errehymy,
``Static and spherically symmetric wormholes in power-law f(R) gravity model,''
Phys. Dark Univ. \textbf{44}, 101438 (2024)
doi:10.1016/j.dark.2024.101438

\bibitem{Shubina:2021tgg}
M.~Shubina,
``Exact analytical vacuum solutions of Rn-gravity model depending on two 
variables,''
Annals Phys. \textbf{451}, 169245 (2023)
doi:10.1016/j.aop.2023.169245
[arXiv:2212.11648 [gr-qc]].


\bibitem{Deng:2014uta}
X.~M.~Deng and Y.~Xie,
``New upper limits on the power of general relativity from solar system dynamics,''
New Astron. \textbf{35}, 36-39 (2014)
doi:10.1016/j.newast.2014.09.003

\bibitem{Ganguly:2013taa}
A.~Ganguly, R.~Gannouji, R.~Goswami and S.~Ray,
``Neutron stars in the Starobinsky model,''
Phys. Rev. D \textbf{89}, no.6, 064019 (2014)
doi:10.1103/PhysRevD.89.064019
[arXiv:1309.3279 [gr-qc]].

\bibitem{Schmidt:2012ms}
H.~J.~Schmidt and D.~Singleton,
``Isotropic universe with almost scale-invariant fourth-order gravity,''
J. Math. Phys. \textbf{54}, 062502 (2013)
doi:10.1063/1.4808255
[arXiv:1212.1769 [gr-qc]].

\bibitem{Jaime:2012yi}
L.~G.~Jaime, L.~Pati{\~n}o and M.~Salgado,
``About matter and dark-energy domination eras in $R^n$ gravity or lack thereof,''
Phys. Rev. D \textbf{87}, no.2, 024029 (2013)
doi:10.1103/PhysRevD.87.024029
[arXiv:1212.2604 [gr-qc]].

\bibitem{DeBenedictis:2012qz}
A.~DeBenedictis and D.~Horvat,
``On Wormhole Throats in $f(R)$ Gravity Theory,''
Gen. Rel. Grav. \textbf{44}, 2711-2744 (2012)
doi:10.1007/s10714-012-1412-x
[arXiv:1111.3704 [gr-qc]].

\bibitem{Gannouji:2011qz}
R.~Gannouji and M.~Sami,
``Vainshtein mechanism in Gauss-Bonnet gravity and Galileon aether,''
Phys. Rev. D \textbf{85}, 024019 (2012)
doi:10.1103/PhysRevD.85.024019
[arXiv:1107.1892 [gr-qc]].

\bibitem{Nzioki:2010nj}
A.~M.~Nzioki, P.~K.~S.~Dunsby, R.~Goswami and S.~Carloni,
``A Geometrical Approach to Strong Gravitational Lensing in f(R) Gravity,''
Phys. Rev. D \textbf{83}, 024030 (2011)
doi:10.1103/PhysRevD.83.024030
[arXiv:1002.2056 [gr-qc]].

\bibitem{Park:2010da}
C.~G.~Park, J.~c.~Hwang and H.~Noh,
``Constraints on a $f(R)$ gravity dark energy model with early scaling evolution,''
JCAP \textbf{09}, 038 (2011)
doi:10.1088/1475-7516/2011/09/038
[arXiv:1012.1662 [astro-ph.CO]].

\bibitem{Leon:2010pu}
G.~Leon and E.~N.~Saridakis,
``Dynamics of the anisotropic Kantowsky-Sachs geometries in $R^n$ gravity,''
Class. Quant. Grav. \textbf{28}, 065008 (2011)
doi:10.1088/0264-9381/28/6/065008
[arXiv:1007.3956 [gr-qc]].

\bibitem{Bisabr:2010sq}
Y.~Bisabr,
``Local Gravity Constraints and Power Law f(R) Theories,''
Grav. Cosmol. \textbf{16}, 239-244 (2010)
doi:10.1134/S0202289310030084
[arXiv:1005.5670 [gr-qc]].

\bibitem{Capozziello:2007vd}
S.~Capozziello, M.~De Laurentis and M.~Francaviglia,
``Higher-order gravity and the cosmological background of gravitational waves,''
Astropart. Phys. \textbf{29}, 125-129 (2008)
doi:10.1016/j.astropartphys.2007.12.001
[arXiv:0712.2980 [gr-qc]].

\bibitem{AvilesCervantes:2008kno}
A.~Aviles Cervantes and J.~L.~Cervantes-Cota,
``Cosmological phase space of $R^n$ gravity,''
AIP Conf. Proc. \textbf{1083}, no.1, 57-64 (2008)
doi:10.1063/1.3058579
[arXiv:0901.3722 [gr-qc]].

\bibitem{Martins:2007uf}
C.~F.~Martins and P.~Salucci,
``Analysis of Rotation Curves in the framework of R**n gravity,''
Mon. Not. Roy. Astron. Soc. \textbf{381}, 1103-1108 (2007)
doi:10.1111/j.1365-2966.2007.12273.x
[arXiv:astro-ph/0703243 [astro-ph]].

\bibitem{Capozziello:2006ph}
S.~Capozziello, V.~F.~Cardone and A.~Troisi,
``Low surface brightness galaxies rotation curves in the low energy limit of r**n 
gravity: no need for dark matter?,''
Mon. Not. Roy. Astron. Soc. \textbf{375}, 1423-1440 (2007)
doi:10.1111/j.1365-2966.2007.11401.x
[arXiv:astro-ph/0603522 [astro-ph]].

\bibitem{Mendoza:2006hs}
S.~Mendoza and Y.~M.~Rosas-Guevara,
``Gravitational waves and lensing of the metric theory proposed by Sobouti,''
Astron. Astrophys. \textbf{472}, 367-371 (2007)
doi:10.1051/0004-6361:20066787
[arXiv:astro-ph/0610390 [astro-ph]].

\bibitem{Sobouti:2006rd} Y.~Sobouti, ``An f(r) gravitation instead of dark 
matter,'' Astron. Astrophys. \textbf{464}, 921 (2007) [erratum: Astron. 
Astrophys. \textbf{472}, 833 (2007)] doi:10.1051/0004-6361:20077452 
[arXiv:0704.3345 [astro-ph]].

\bibitem{CTT} Capozziello, S. Carloni, and A. Troisi, Res. Dev. Astron. Astrophys. 
1, 625 (2003).

\bibitem{Capozziello:2002rd}
S.~Capozziello,
``Curvature quintessence,''
Int. J. Mod. Phys. D \textbf{11}, 483-492 (2002)
doi:10.1142/S0218271802002025
[arXiv:gr-qc/0201033 [gr-qc]].

\bibitem{Capozziello:2009jg}
S.~Capozziello, M.~De Laurentis and A.~Stabile,
``Axially symmetric solutions in f(R)-gravity,''
Class. Quant. Grav. \textbf{27}, 165008 (2010)
doi:10.1088/0264-9381/27/16/165008
[arXiv:0912.5286 [gr-qc]].

\bibitem{Dunsby:2009zz}
P.~K.~S.~Dunsby,
``The evolution of density fluctuations in modified theories of gravity,''
AIP Conf. Proc. \textbf{1115}, no.1, 205-211 (2009)
doi:10.1063/1.3131500

\bibitem{Faraoni:2009xb}
V.~Faraoni,
``Clifton's spherical solution in f(R) vacuo harbours a naked singularity,''
Class. Quant. Grav. \textbf{26}, 195013 (2009)
doi:10.1088/0264-9381/26/19/195013
[arXiv:0909.0514 [gr-qc]].

\bibitem{Goheer:2009ss}
N.~Goheer, J.~Larena and P.~K.~S.~Dunsby,
``Power-law cosmic expansion in f(R) gravity models,''
Phys. Rev. D \textbf{80}, 061301 (2009)
doi:10.1103/PhysRevD.80.061301
[arXiv:0906.3860 [gr-qc]].

\bibitem{Goheer:2008tn}
N.~Goheer, R.~Goswami and P.~K.~S.~Dunsby,
``Dynamics of f(R)-cosmologies containing Einstein static models,''
Class. Quant. Grav. \textbf{26}, 105003 (2009)
doi:10.1088/0264-9381/26/10/105003
[arXiv:0809.5247 [gr-qc]].

\bibitem{Ananda:2008tx}
K.~Ananda, S.~Carloni and P.~K.~S.~Dunsby,
``A detailed analysis of structure growth in $f(R)$ theories of gravity,''
Class. Quant. Grav. \textbf{26}, 235018 (2009)
doi:10.1088/0264-9381/26/23/235018
[arXiv:0809.3673 [astro-ph]].

\bibitem{Carloni:2007br}
S.~Carloni, A.~Troisi and P.~K.~S.~Dunsby,
``Some remarks on the dynamical systems approach to fourth order gravity,''
Gen. Rel. Grav. \textbf{41}, 1757-1776 (2009)
doi:10.1007/s10714-008-0747-9
[arXiv:0706.0452 [gr-qc]].

\bibitem{Carloni:2008jy}
S.~Carloni, K.~N.~Ananda, P.~K.~S.~Dunsby and M.~E.~S.~Abdelwahab,
``Unifying the study of background dynamics and perturbations in $f(R)$-gravity,''
[arXiv:0812.2211 [astro-ph]].

\bibitem{Ananda:2008gs}
K.~N.~Ananda, S.~Carloni and P.~K.~S.~Dunsby,
``A characteristic signature of fourth order gravity,''
Springer Proc. Phys. \textbf{137}, 165-172 (2011)
doi:10.1007/978-3-642-19760-4{\_}15
[arXiv:0812.2028 [astro-ph]].

\bibitem{Goheer:2007wx}
N.~Goheer, J.~A.~Leach and P.~K.~S.~Dunsby,
``Compactifying the state space for alternative theories of gravity,''
Class. Quant. Grav. \textbf{25}, 035013 (2008)
doi:10.1088/0264-9381/25/3/035013
[arXiv:0710.0819 [gr-qc]].

\bibitem{Ananda:2007xh}
K.~N.~Ananda, S.~Carloni and P.~K.~S.~Dunsby,
``The Evolution of cosmological gravitational waves in f(R) gravity,''
Phys. Rev. D \textbf{77}, 024033 (2008)
doi:10.1103/PhysRevD.77.024033
[arXiv:0708.2258 [gr-qc]].

\bibitem{Carloni:2007yv}
S.~Carloni, P.~K.~S.~Dunsby and A.~Troisi,
``The Evolution of density perturbations in f(R) gravity,''
Phys. Rev. D \textbf{77}, 024024 (2008)
doi:10.1103/PhysRevD.77.024024
[arXiv:0707.0106 [gr-qc]].

\bibitem{Amendola:2006we}
L.~Amendola, R.~Gannouji, D.~Polarski and S.~Tsujikawa,
``Conditions for the cosmological viability of f(R) dark energy models,''
Phys. Rev. D \textbf{75}, 083504 (2007)
doi:10.1103/PhysRevD.75.083504
[arXiv:gr-qc/0612180 [gr-qc]].

\bibitem{Goheer:2007wu}
N.~Goheer, J.~A.~Leach and P.~K.~S.~Dunsby,
``Dynamical systems analysis of anisotropic cosmologies in 
$R^n$-gravity,''
Class. Quant. Grav. \textbf{24}, 5689-5708 (2007)
doi:10.1088/0264-9381/24/22/026
[arXiv:0710.0814 [gr-qc]].

\bibitem{Clifton:2007ih}
T.~Clifton,
``Exact Friedmann Solutions in Higher-Order Gravity Theories,''
Class. Quant. Grav. \textbf{24}, 5073-5091 (2007)
doi:10.1088/0264-9381/24/20/010
[arXiv:gr-qc/0703126 [gr-qc]].
	
\bibitem{Leach:2007ss}
J.~A.~Leach, P.~K.~S.~Dunsby and S.~Carloni,
``An Analysis of the shear dynamics in Bianchi I cosmologies with R**n-gravity,''
doi:10.1142/9789812834300{\_}0109
[arXiv:gr-qc/0702122 [gr-qc]].

\bibitem{Carloni:2006mr}
S.~Carloni and P.~K.~S.~Dunsby,
``A Dynamical system approach to higher order gravity,''
J. Phys. A \textbf{40}, 6919-6926 (2007)
doi:10.1088/1751-8113/40/25/S40
[arXiv:gr-qc/0611122 [gr-qc]].

\bibitem{Clifton:2006kc}
T.~Clifton and J.~D.~Barrow,
``Further exact cosmological solutions to higher-order gravity theories,''
Class. Quant. Grav. \textbf{23}, 2951 (2006)
doi:10.1088/0264-9381/23/9/011
[arXiv:gr-qc/0601118 [gr-qc]].

\bibitem{Carloni:2005ii}
S.~Carloni, P.~K.~S.~Dunsby and D.~M.~Solomons,
``Bounce conditions in f(R) cosmologies,''
Class. Quant. Grav. \textbf{23}, 1913-1922 (2006)
doi:10.1088/0264-9381/23/6/006
[arXiv:gr-qc/0510130 [gr-qc]].

\bibitem{Capozziello:2006dp}
S.~Capozziello, V.~F.~Cardone and A.~Troisi,
``Gravitational lensing in fourth order gravity,''
Phys. Rev. D \textbf{73}, 104019 (2006)
doi:10.1103/PhysRevD.73.104019
[arXiv:astro-ph/0604435 [astro-ph]].

\bibitem{Leach:2006br}
J.~A.~Leach, S.~Carloni and P.~K.~S.~Dunsby,
``Shear dynamics in Bianchi I cosmologies with R**n-gravity,''
Class. Quant. Grav. \textbf{23}, 4915-4937 (2006)
doi:10.1088/0264-9381/23/15/011
[arXiv:gr-qc/0603012 [gr-qc]].

\bibitem{Clifton:2006ug}
T.~Clifton,
``Spherically Symmetric Solutions to Fourth-Order Theories of Gravity,''
Class. Quant. Grav. \textbf{23}, 7445 (2006)
doi:10.1088/0264-9381/23/24/015
[arXiv:gr-qc/0607096 [gr-qc]].

\bibitem{Carloni:2004kp}
S.~Carloni, P.~K.~S.~Dunsby, S.~Capozziello and A.~Troisi,
``Cosmological dynamics of R**n gravity,''
Class. Quant. Grav. \textbf{22}, 4839-4868 (2005)
doi:10.1088/0264-9381/22/22/011
[arXiv:gr-qc/0410046 [gr-qc]].

\bibitem{Clifton:2005aj}
T.~Clifton and J.~D.~Barrow,
``The Power of General Relativity,''
Phys. Rev. D \textbf{72}, no.10, 103005 (2005)
[erratum: Phys. Rev. D \textbf{90}, no.2, 029902 (2014)]
doi:10.1103/PhysRevD.72.103005
[arXiv:gr-qc/0509059 [gr-qc]].

\bibitem{Furey:2004rq}
N.~Furey and A.~DeBenedictis,
``Wormhole throats in R**m gravity,''
Class. Quant. Grav. \textbf{22}, 313-322 (2005)
doi:10.1088/0264-9381/22/2/005
[arXiv:gr-qc/0410088 [gr-qc]].

\bibitem{Pavlov:1997xf}
A.~Pavlov,
``Two-dimensional R**n gravitation,''
Int. J. Theor. Phys. \textbf{36}, 2107-2113 (1997)
doi:10.1007/BF02435947

\bibitem{Ferraris:1992dx}
M.~Ferraris, M.~Francaviglia and I.~Volovich,
``The Universality of vacuum Einstein equations with cosmological constant,''
Class. Quant. Grav. \textbf{11}, 1505-1517 (1994)
doi:10.1088/0264-9381/11/6/015
[arXiv:gr-qc/9303007 [gr-qc]].

\bibitem{Ciftci:2017tjc}
D.~K.~{\c{C}}iftci and V.~Faraoni,
``Perfect fluid solutions of Brans{\textendash}Dicke and $f(R)$ cosmology,''
Annals Phys. \textbf{391}, 65-82 (2018)
doi:10.1016/j.aop.2018.02.002
[arXiv:1711.04026 [gr-qc]].

\bibitem{Nguyen:2023whv} H.~K.~Nguyen, ``Emerging Newtonian potential in 
pure R$^{2}$ gravity on a de Sitter background,'' JHEP \textbf{08}, 127 
(2023) doi:10.1007/JHEP08(2023)127 [arXiv:2306.03790 [gr-qc]].

\bibitem{Barker:2025gon}
W.~Barker and D.~Glavan,
``Spectrum of pure $R^2$ gravity: full Hamiltonian analysis,''
[arXiv:2510.08201 [gr-qc]].

\bibitem{Hell:2023mph}
A.~Hell, D.~L\"{u}st and G.~Zoupanos,
``On the degrees of freedom of R$^{2}$ gravity in flat spacetime,''
JHEP \textbf{02}, 039 (2024)
doi:10.1007/JHEP02(2024)039
[arXiv:2311.08216 [hep-th]].

\bibitem{Hell:2025wha}
A.~Hell and D.~L\"{u}st,
``Conformal and pure scale-invariant gravities in d dimensions,''
JHEP \textbf{09}, 202 (2025)
doi:10.1007/JHEP09(2025)202
[arXiv:2506.18775 [hep-th]].

\bibitem{Hell:2025lbl}
A.~Hell and D.~L\"{u}st,
``Aspects of non-minimally coupled curvature with power laws,''
[arXiv:2509.20217 [hep-th]].

\bibitem{DeWitt65} B. De Witt, {\it Dynamical Theory of Groups and Fields} 
(Gordon and Breach, New York, 1965).

\bibitem{Faraoni:2022doe}
V.~Faraoni, A.~Giusti, S.~Jose and S.~Giardino,
``Peculiar thermal states in the first-order thermodynamics of gravity,''
Phys. Rev. D \textbf{106}, no.2, 024049 (2022)
doi:10.1103/PhysRevD.106.024049
[arXiv:2206.02046 [gr-qc]].

\bibitem{Faraoni:2022fxo}
V.~Faraoni, P.~A.~Graham and A.~Leblanc,
``Critical solutions of nonminimally coupled scalar field theory and 
first-order thermodynamics of gravity,''
Phys. Rev. D \textbf{106}, no.8, 084008 (2022)
doi:10.1103/PhysRevD.106.084008
[arXiv:2207.03841 [gr-qc]].

\bibitem{Faraoni:2022jyd}
V.~Faraoni and T.~B.~Fran\c{c}onnet,
``Stealth metastable state of scalar-tensor thermodynamics,''
Phys. Rev. D \textbf{105}, no.10, 104006 (2022)
doi:10.1103/PhysRevD.105.104006
[arXiv:2203.14934 [gr-qc]].

\bibitem{NariaiTomita1971} H. Nariai and K. Tomita, ``On the removal of 
initial singularity in a Big-Bang universe in terms of a renormalized 
theory of gravitation. II: criteria for obtaining a physically reasonable 
model'', {\it Prog. Theor. Phys.} \textbf{46} (3), 776-786 (1971).

\bibitem{Giesswein:1974fog}
M.~Giesswein, R.~Sexl and E.~Streeruwitz,
``Cosmological singularities and higher-order gravitational 
lagrangians,''
Phys. Lett. B \textbf{52}, 442-444 (1974)
doi:10.1016/0370-2693(74)90120-8

\bibitem{Faraoni:2005vk}
V.~Faraoni and S. Nadeau,
``The Stability of modified gravity models,''
Phys. Rev. D \textbf{72}, 124005 (2005)
doi:10.1103/PhysRevD.72.124005
[arXiv:gr-qc/0511094 [gr-qc]].

\end{thebibliography}
\end{document}